\documentstyle[11pt,aaspp4]{article}   
 
\slugcomment{Submitted to the {\it Astronomical Journal}}
 
\lefthead{Lauer et al.}
\righthead{${\rm M}32\pm1$}
 
\def\asec{\rlap{$^{\prime\prime}$}.\hbox to 2pt{}}
 
\begin{document}

\title{${\rm M}32\pm1$\altaffilmark{1}\altaffilmark{2}}

\author{Tod R. Lauer} 
\affil{National Optical Astronomy
Observatories\altaffilmark{3}, P.O. Box 26732, Tucson, Arizona 85726}
\affil{Electronic mail: lauer@noao.edu}

\author{S. M. Faber}
\affil{UCO/Lick Observatory, Board of Studies in Astronomy and
Astrophysics, University of California, Santa Cruz, California 95064}
\affil{Electronic mail: faber@ucolick.org}

\author{Edward A. Ajhar} 
\affil{National Optical Astronomy
Observatories\altaffilmark{3}, P.O. Box 26732, Tucson, Arizona 85726}
\affil{Electronic mail: ajhar@noao.edu}

\author{Carl J. Grillmair}
\affil{SIRTF Science Center, California Institute of Technology,
770 S. Wilson, Pasadena, California 91125}
\affil{Electronic mail: carl@bb1.jpl.nasa.gov}

\and

\author{Paul A. Scowen}
\affil{Arizona State University, Tempe, Arizona, 85287}
\affil{Electronic mail: paul.scowen@asu.edu}

\altaffiltext{1}{We wish to dedicate this paper to the memory
of Robert M. Light, whose selfless service to the WFPC-1 IDT
made this work and many others possible.}
\altaffiltext{2}{Based on observations with the NASA/ESA {\it Hubble
Space Telescope,} obtained at the Space Telescope Science Institute
(STScI), which is operated by the Association of Universities for
Research in Astronomy (AURA), Inc., under National Aeronautics and
Space Administration (NASA) Contract NAS 5-26555.}
\altaffiltext{3}{The National Optical Astronomy Observatories are
operated by AURA, Inc., under cooperative agreement with the National
Science Foundation.}

\begin{abstract}

Multicolor {\it HST} WFPC-2 images are used to study the central
structure of the three Local Group galaxies M31, M32, and M33.
PSF-deconvolution and modeling of image aliasing are
required to recover accurate brightness profiles within $r<0\asec5.$\@
The data present a study in contrasts that suggests
different evolutionary histories.
{\bf In M31,}  the nucleus is double-peaked, as found by WFPC-1 and
confirmed by WFPC-2.
The dimmer peak, P2, is closely centered on the bulge isophotes to
$0\asec1$, implying that it is the dynamical center of the galaxy.
Directly on P2 lies a UV-bright compact source
that was discovered by King et al.\ (1995) at 1700 \AA.\@
WFPC-2 images now show that this source is {\it resolved,}
with  $r_{1/2}\approx0.2~{\rm pc}.$\@
It dominates the nucleus at 3000 \AA,
and its spectral energy distribution is consistent with late B-early A stars.
This probable nuclear star cluster may consist of young stars
and be an older version of the central
cluster of hot stars that now sits at the center of the Milky Way, 
or it may consist of heavier stars built up from collisions in a
possible cold disk of stars orbiting P2.
Aspects of its formation remain highly problematic.
{\bf In M32,} new images show that the central
cusp continues to rise into the {\it HST} limit with slope
$\gamma\approx0.5,$ and the central density $\rho_0>10^7M_\odot~{\rm pc}^{-3}.$\@
The $V-I$ and $U-V$ color profiles are essentially flat, and
there is no sign of an inner disk, dust, or any other structure.
This total lack of features seems at variance with a nominal stellar collision
time of $2 \times 10^{10}$ yr, which implies that a significant
fraction of the light in the central pixel should come from blue stragglers.
This discrepancy is eased but not completely
removed if the stellar population is young ($2 \times 10^9$ yr).  
The stubborn normalcy of M32 at tiny radii may be emerging as
an important puzzle.
{\bf In M33,} the nucleus has an extremely steep $\gamma=1.49$ power-law
profile for $0\asec05<r<0\asec2$ that appears to become somewhat
shallower as the {\it HST} resolution limit is approached.
The profile for $r<0\asec04$ can be described as having either a
$\gamma\approx0.8$ cusp or a small core with $r_c\approx0.13~{\rm pc}.$\@
The central density is $\rho_0>2\times10^6M_\odot~{\rm pc}^{-3}$, and
the implied central relaxation time is only $\sim3\times10^6$ yr,
indicating that the nucleus is highly relaxed.
The accompanying short collision time of $7\times 10^9$ yr predicts a central
blue straggler component that is quantitatively consistent with the strong
$V-I$ and $B-R$ color gradients seen with {\it HST} and from the ground.
When combined with the Galaxy, the nuclei of the Local Group
show surprisingly similar radial density profiles but
divide into two classes according to velocity dispersion
and black hole content:  M31, M32, and the Milky Way
are dominated dynamically (and stabilized against relaxation) by massive
central black holes, while M33 lacks a dominant black hole. 
An obvious hypothesis is that a sizable stellar spheroid
(which M33 lacks) is necessary to grow a massive black hole.
A further implication is that the black hole growth in M31, M32,
and the Milky Way was accompanied by evolution in the stellar density profiles,
stellar populations, and dynamical structure of these nuclei such that their
past appearance may have differed significantly from what they look like today.
In short, {\it HST} observations are taking us to scales where understanding
the central structure of galactic nuclei is intimately connected to
the detailed interactions among their central stellar populations.

\end{abstract}


\section{Introduction}

The three Local Group galaxies M31, M32, and M33 have central structures
as diverse as their global morphologies.
M31 has a double nucleus of unknown origin embedded in a
bulge that has a relatively large core
(\markcite{l93}Lauer et al.\ 1993).
M32 has a smooth brightness profile that monotonically rises to extreme
stellar densities but that contains no separate nuclear component
(\markcite{l92}Lauer et al.\ 1992).
M33, in contrast, has an extremely compact star cluster at its center
that has little apparent relation to the surrounding galaxy.
M31 and M32 both have strong nuclear rotation and velocity
dispersion gradients that are evidence for massive central black holes
(for M31 see \markcite{dr88}Dressler \& Richstone 1988,
\markcite{k88}Kormendy 1988, and \markcite{bac}Bacon et al.\ 1994;
for M32 see \markcite{ton}Tonry 1984, \markcite{ton2}Tonry 1987,
\markcite{dr88}Dressler \& Richstone 1988, \markcite{ben96}Bender,
Kormendy, \& Dehnen 1996,
\markcite{vdm97}van der Marel et al.\ 1997,
and \markcite{vdm98}van der Marel et al.\ 1998),
while the low rotation and velocity dispersion of the M33 nucleus
place tight upper limits on the mass of any central 
dark object (\markcite{km}Kormendy \& McClure 1993).

One common element among these galaxies is their proximity.
{\it HST} observations of all three systems have
spatial resolution of only $\sim0.2~{\rm pc,}$ a scale inaccessible
for any other galactic system beyond the Local Group.
Sub-parsec resolution proves crucial for even beginning
to reveal the central properties of these galaxies.
M31, M32, and M33 thus may be important ``stand-ins'' for galaxies
at much larger distances.
Extensive work on M31 and M32 
was conducted with WFPC-1 and FOC 
prior to the spherical aberration correction.
WFPC-2 observations now offer the opportunity to
revisit these galaxies with both improved spatial resolution and
dynamic range, with the hope of clarifying some of the issues
unresolved by earlier work.

Prior to {\it HST,} the central structure of M31 was
considered to be the prototypical example of a dense nuclear star
cluster embedded within the core of a separate bulge system.
{\it HST,} however, showed that the M31 nucleus was double.
The light that had dominated the ground-based images and spectra
at optical and near-IR wavelengths actually came from a bright component
(designated P1 by \markcite{l93}Lauer et al.\
1993) apparently physically separate and displaced by $\sim0\asec5$
from a less luminous system, P2, that itself was at the true photometric
and dynamical center of the bulge.
One obvious explanation was that P1 represented the final stages of
another galaxy being cannibalized by M31; however, this picture
presented numerous difficulties related to the timing, dynamics,
and detailed physical properties of the
central components (\markcite{l93}Lauer et al.\ 1993).
\markcite{trem}Tremaine (1995) suggested that the brightness peak at P1 could
be explained as the apogalatica of stars orbiting P2 in an eccentric disk.
Separately, \markcite{k95}King, Stanford, \& Crane (1995) showed
that P2 actually outshone P1 at far-UV wavelengths,
with perhaps some of the light associated with P2 having a nonthermal origin.

The resolution of the WFPC-1 and FOC was more than adequate
to reveal the important morphological details of the M31 nucleus.
Understanding its complex form, and testing the Tremaine
model in particular, is more likely to require
higher-resolution dynamics than better imaging data.
Nevertheless, the new WFPC-2 observations presented here have vastly
improved dynamical range over the older data, permitting careful
tests for color gradients within the nucleus
that might indicate interesting changes in the stellar population,
as well as allowing a search for subtle effects of dust or other components.
The data have been also taken with extremely high $S/N$ and improved
spatial sampling to reveal any secular morphological evolution
of the nucleus at future epochs.

The central structure of M32 may be typical of small elliptical
galaxies in general.
WFPC-1 showed that the central stellar density of M32 may exceed
$10^7M_\odot~{\rm pc}^{-3}$ for $r<0.2~{\rm pc,}$,
yet its density profile at $r\sim5~{\rm pc,}$ a scale accessible with
{\it HST} in the Virgo cluster, is typical
for a low-luminosity elliptical (\markcite{l95}Lauer et al.\ 1995).
Further, M32 shows no evidence for harboring a separate nuclear star
cluster, disk, dust, or any other structure apart from the smooth,
monotonic rise of the overall stellar density distribution.
A remaining issue
is the precise form of the central profile
at $r<0.4~{\rm pc,}$ which is poorly constrained by the
WFPC-1 images.  \markcite{l92}Lauer et al.\ (1992) showed that M32 could be
fitted with models that ranged from those with a small $r_c=0.37~{\rm pc}$ core,
to those with a $\gamma\approx0.5$ cusp interior to this radius,
with a corresponding range in central density of
$4\times10^6M_\odot~{\rm pc}^{-3}$ to $3\times10^7M_\odot~{\rm pc}^{-3}.$\@
This scale is of sharp interest as central relaxation and
even stellar collisions are important over the
age of the universe at these densities,
unless the central dynamics are strongly dominated by a massive black hole.
The new WFPC-2 observations of M32 have $\sim2\times$ the previous
spatial resolution, allowing for a much improved {\it lower} limit
to the central density of M32.
Further, as for M31, the improved dynamic range allows for the
detection of subtle color gradients that may be associated with
changes in the central stellar population.

The nuclear star cluster at the center of M33 may reach central stellar
densities similar to that of M32 (\markcite{km}Kormendy \& McClure 1993).
Kormendy \& McClure
further showed that the nucleus has a strong $B-R$ color gradient in the
sense that the center of the cluster is {\it bluer} than its outskirts.
They discuss the intriguing possibility that stellar collisions
at the center of the M33 nucleus have substantially altered its
stellar population by making a young generation of blue stragglers.
The contrast with M32 is particularly interesting.
\markcite{l92}Lauer et al.\ (1992) argue
from the lack of color gradients and other data that relaxation and stellar
collisions have not taken place in the center of M32.
The cross-sections for relaxation and physical collisions between stars
decrease rapidly with increasing stellar velocity dispersion
(for velocities less than the stellar escape velocities), due to
the decreasing size of the gravitational sphere of influence
of the individual stars.
The $\sim3\times10^6M_\odot$ black hole believed to be at the center of M32
(see \markcite{kr}Kormendy \& Richstone 1995 for a review)
would thus raise orbital speeds and greatly lengthen
the relaxation and collision time scales in M32.
Kormendy \& McClure, in contrast, show that M33 has a central velocity
dispersion of only 21 km s$^{-1}$, thus making it highly likely that stellar
relaxation and physical collisions have strongly modified the
structure of the M33 nucleus.
A basic goal of the new WFPC-2 observations is to obtain an improved
estimate of the central density of the M33 cluster, as well as to
explore further the centrally blue color gradient detected by
Kormendy \& McClure.

We present the new images and profiles of M31, M32, and M33,
followed by comparisons and contrasts among the three galaxies
and the Milky Way nucleus in the following sections.
Our overall conclusion is that M31, M32, and the Milky Way
have had similar formation histories but that
M33 stands apart owing to its lack of a black hole.  This
difference may be related to the presence of a spheroidal
component in the first three galaxies but the lack of one in M33.
However, even the black-hole nuclei differ in detail,
and the differences beg an explanation.
An important part of this work is understanding
the resolution limits of the new WFPC-2 data.
Although the spherical aberration correction in the WFPC-2 optics
represents a vast improvement over the aberrated WFPC-1 observations,
the {\it HST}+WFPC-2 PSF remains important, and its blurring effects
must be accounted for in quantitative work by either
deconvolution or modeling.

\vskip\baselineskip
\vskip\baselineskip

\section{Observations and Analysis \label{sec:obs} }

\subsection{Observational Material}
 
The WFPC-2 observations of M31, M32, and M33 are summarized
in Table \ref{tab:obs}.
The galaxies were centered in the high-resolution PC1 CCD for all images.
The exposure times were set with the general goal of exceeding
$S/N\sim100$ in the central pixel of the galaxies, although
time constraints prevented the F300W exposures of M31 from achieving this level.
In addition, little if any galaxy light was seen in the F160W exposures
of M31 despite the long exposure, due to the
poor far-UV sensitivity of WFPC-2.
A low gain setting ($15e^-/$DN) was used for the F555W and F814W
exposures to reach $S/N$ levels high enough to measure
the separation of P1 and P2 with especially high precision.
The $V$ and $I$ M31 exposures were also dithered by 0.5 PC pixels, mapping out
a $2\times2$ subpixel pattern to counter the undersampling present in PC images.
Dithered F555W images of the quadrupole gravitational lens 2237+0305
(\markcite{huchra}discovered by Huchra et al.\ 1985) were also obtained
near in time to M31,
with the lens positioned at the same location on the PC1 CCD.
These observations are not presented here but can serve as an astrometric
reference for M31 observations taken at future epochs.
Deep exposures at the normal gain setting ($7e^-/$DN) were also obtained
in F555W and F814W for M31, and F555W for M33, to look for faint point sources
near the centers of these galaxies.

Basic reduction of the images was accomplished with calibration
products kindly provided by the WFPC-2 Investigation Definition Team (IDT).
The observations were obtained with the telescope guiding in
fine-lock, and, with the exception of the dithered F555W and F814W exposures
of M31, the multiple images in each filter sequence were generally
registered to better than 0.05 pixels, enabling excellent cosmic
ray rejection.
The guiding performance for all images was excellent, with the jitter always
being $0\asec004$ or less.

Cosmic ray rejection was complex for the dithered M31 images,
as the 0.5 pixel shifts were enough to cause
significant changes in the image structure that could be mistaken
for cosmic ray events by a simple statistical examination of the pixel values.
The rejection of cosmic rays was done iteratively;
cosmic rays in a given image were identified by using
an average of the remaining frames shifted to its position for comparison.
This process was repeated a second time after initial identification
of cosmic rays in the comparison frames.

The final F555W and F814W images of M31 were built from the dithered
sets by simply interleaving the four images for each filter.
The images used for analysis thus have pixels spaced every
$0\asec0228,$ or 0.5 PC pixel.
These images can be viewed as M31 convolved with the HST+PC PSF
{\it and} a box the size of a PC pixel, but then sampled with
a grid $2\times$ finer than that of the PC.
The F555W and F814W M31 images are thus Nyquist-sampled.

\subsection{Image Deconvolution\label{sec:decon}}

\begin{figure}[bhtp]
\epsscale{0.6}
\plotone{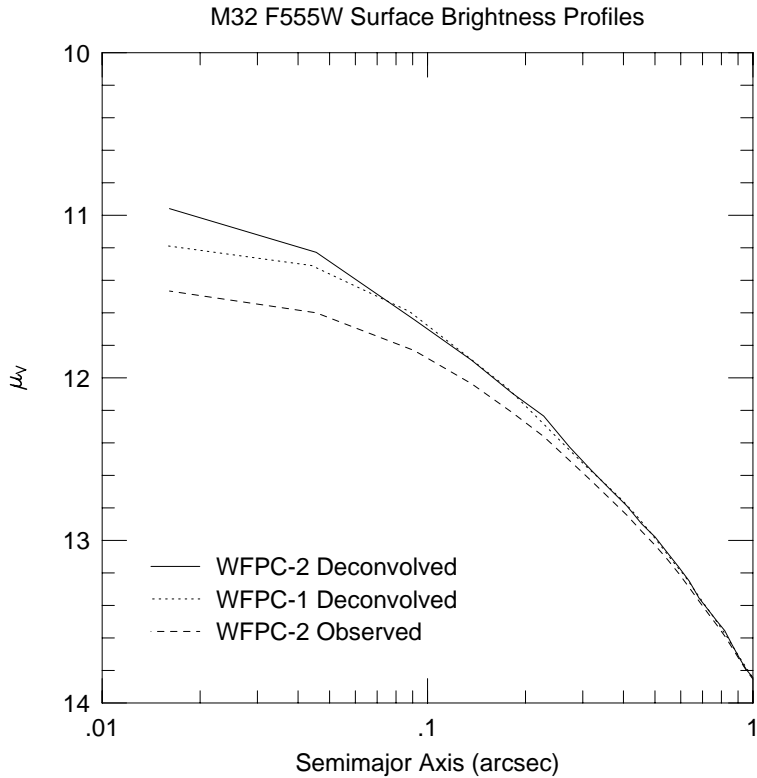}
\caption{The inner $1''$ of the WFPC-2 deconvolved {\it V} surface
brightness profile of M32 compared to the WFPC-2 profile measured prior to
deconvolution and the deconvolved WFPC-1 profile (Lauer et al.\ 1992).
Note that the deconvolved WFPC-1 profile is more accurate for $r\geq0\asec1$
than the WFPC-2 profile {\it prior} to deconvolution.}
\label{fig:m32_pro}
\end{figure}

\begin{figure}[ht]
\epsscale{0.6}
\plotone{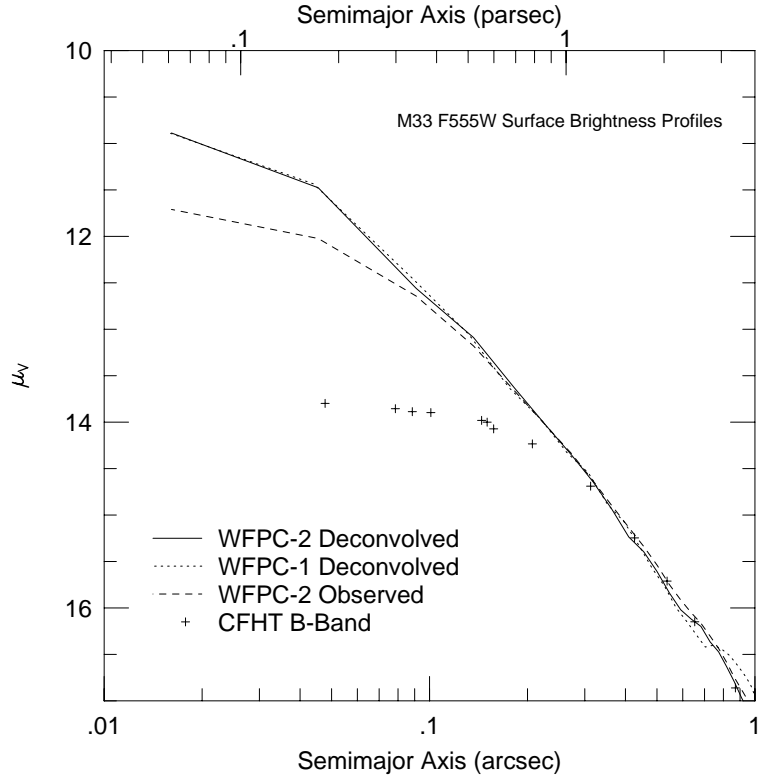}
\caption{The inner $1''$ of the WFPC-2 deconvolved {\it V} surface
brightness profile of M33 compared to the profile measured prior to
deconvolution, and the deconvolved WFPC-1 profile.
Again note the excellent agreement of the WFPC-1 and WFPC-2 deconvolved
profiles.  The CHFT B1 ({\it B}-band) profile of Kormendy \& McClure (1993)
is also shown for comparison, assuming $B-V=0.45.$\@}
\label{fig:m33_pro}
\end{figure}

Despite the dramatic improvement in image quality provided by
the corrective optics installed in WFPC-2,
PSF blurring is still important in the new data,
and deconvolution or modeling of the images will still be
required for quantitative analysis.
This can be seen clearly in Figures \ref{fig:m32_pro}
and \ref{fig:m33_pro}, which compare the F555W
brightness profiles for M32 and M33
before and after 40 iterations of Lucy-Richardson deconvolution
(\markcite{rich}Richardson 1972; \markcite{lucy}Lucy 1974)
to the deconvolved WFPC-1 profiles of the same systems.
The central few pixels of the profiles as observed with WFPC-2 are clearly
depressed compared to the same profiles after deconvolution.
Indeed, in M32 the effects of blurring persist
at greater than 5\%\ of the local surface brightness as
far out as $0\asec5.$\@
In contrast, the deconvolved profiles of both WFPC-1 and WFPC-2
are in excellent agreement at the few percent level
for $r>0\asec1,$ with the superior resolution of WFPC-2
evident in the central two points in the profiles.
Deconvolution simulations of WFPC-1 data conducted by
\markcite{l92}Lauer et al.\
(1992) had predicted that the WFPC-1 data could provide accurate
brightness profiles for $r>0\asec1$ after deconvolution;
the current deconvolved WFPC-2 profiles provide direct
verification of this.
Indeed, as can be seen in both Figures \ref{fig:m32_pro}
and \ref{fig:m33_pro},
the WFPC-1 deconvolved profiles actually prove more faithful to the brightness
distributions of the galaxies than do the WFPC-2 profiles prior to
deconvolution.

\begin{figure}[ht]
\epsscale{0.6}
\plotone{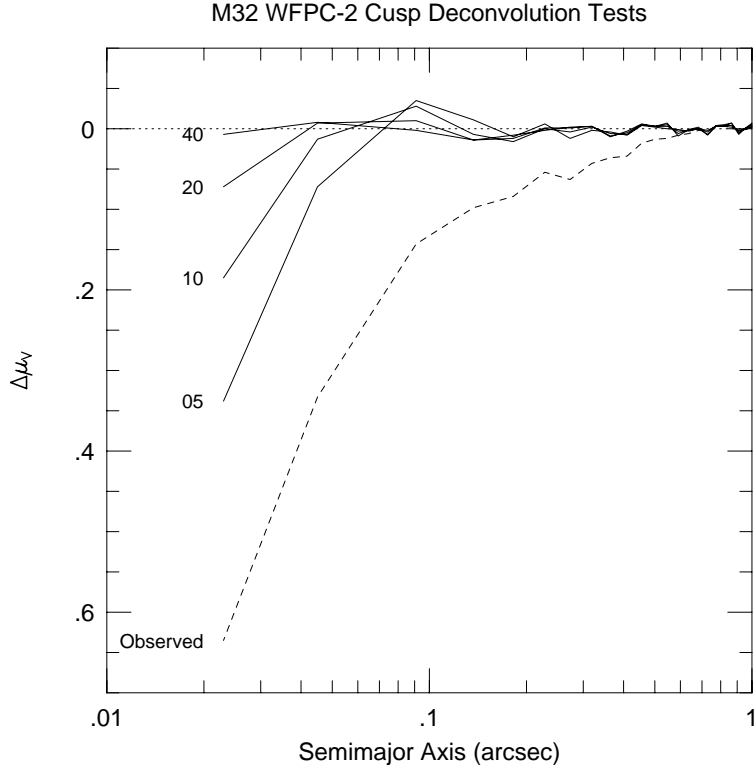}
\caption{Surface brightness comparisons between the Lauer et al.\
(1992) cusp model of M32 and observed and deconvolved profiles.
The dashed line shows the difference between the model and
the profile as observed in a WFPC-2 image after convolution
with the PSF.  Note that while the largest differences
occur for $r<0\asec1,$ which approaches the scale of the PSF core,
blurring effects are still visible at $r\sim0\asec5.$\@
The solid traces show how the profile converges to the model as a function
of the number of Lucy-Richardson deconvolution iterations.}
\label{fig:deconsim}
\end{figure}

To explore the effects of deconvolution further,
deconvolution simulations are presented in Figure \ref{fig:deconsim} on the
same M32 galaxy model simulated by \markcite{l92}Lauer et al.\ (1992)
but with $S/N$ appropriate to the present observations.
The figure shows the residuals of profiles at various levels
of deconvolution as compared to the input model.
The dashed line shows the difference between
the model and observation prior to deconvolution.
verifying the difference between the WFPC-2 observed and deconvolved
profiles of M32 shown in Figure \ref{fig:m32_pro}.
As can be seen, most of the deconvolution correction takes
place in the first few iterations of Lucy-Richardson.
Ten iterations
restore the profile to about the
same level of resolution as did 80 iterations on WFPC-1 images.
After 40 iterations, the profile matches the model with errors
of typically only a few percent over the entire profile;
for $r\geq0\asec04$ (outside the central pixel),
the deconvolved profiles can be taken
as a faithful rendition of the intrinsic galaxy profile to $\sim1$\%.
In this particular experiment,
even the central pixel is recovered at the few percent level.
As was emphasized by Lauer et al., however, the exact central deficit
depends on the input model; the particular residual in this
experiment should not be applied as a general correction.
Again, image modeling is likely to be required at this stage.
The final residual blurring of the central pixel is also
considerably smaller in WFPC-2 than in WFPC-1, allowing the possibility of
using an integrated central luminosity constraint, which was
essentially impossible with WFPC-1.
 
Evaluating the deconvolved profiles also requires understanding the effects
of uncertainties in the PSF.
We evaluate this potential source of systematic error
by using different stellar images as representations of the PSF.
Fortunately, it appears that plausible variations in the PSF produce
both clear and benign effects in the deconvolved profile.
The sole effect of plausible uncertainties in the
core of the PSF is to modulate the brightness of the central pixel
relative to its immediate neighbors.
In particular, we find that the central pixel may vary by $0.06$ mag
when deconvolved with different PSFs.
A more important problem is how the galaxy images are centered within a pixel,
as at F555W the WFPC-2 images are undersampled.
Such undersampling causes aliasing, which can artificially
broaden sharp cusps or nuclei at the galaxy centers.
We discuss the particular effects of this error on M32 and M33 below;
but, in brief, after accounting for aliasing one can
discriminate among various interesting possibilities for
the intrinsic central brightnesses of the two galaxies.
A final concern is allowing for the extended wings of the PSF.
The wings only contribute to low wavenumbers in the Fourier
domain; thus they mainly affect the accuracy of the overall
slope of the profile on arcsecond scales rather than
at the {\it HST} resolution limit.
Nevertheless, because color gradients are generally very shallow in galaxies,
accurate representation of the PSF wings in all colors is vital
for measuring such gradients.

The PSFs used for the deconvolutions were
constructed from several individual star exposures
obtained over the time span of the galaxy observations.
The bulk of the images came from a routine monthly calibration program
that imaged a standard star centered in the PC.
Other data came from WFPC-1 IDT stellar photometry programs.
For the F555W PSF, the dithered gravitational lens images
of 2237+0305 provided excellent sampling of the PSF core.
In all cases, only deep exposures that had well-exposed wings were used.
The composite PSFs have total exposure levels of $\sim10^6e^-$
and extend out to $\sim3'';$ the size of the images is
$256\times256$ pixels, or $11\asec6\times11\asec6$
(the PSFs used for the dithered M31 images have the same
angular extent but double the sampling).

\subsection{M31\label{sec:om31}}
 
\begin{figure}[htbp]
\epsscale{0.5}
\plotone{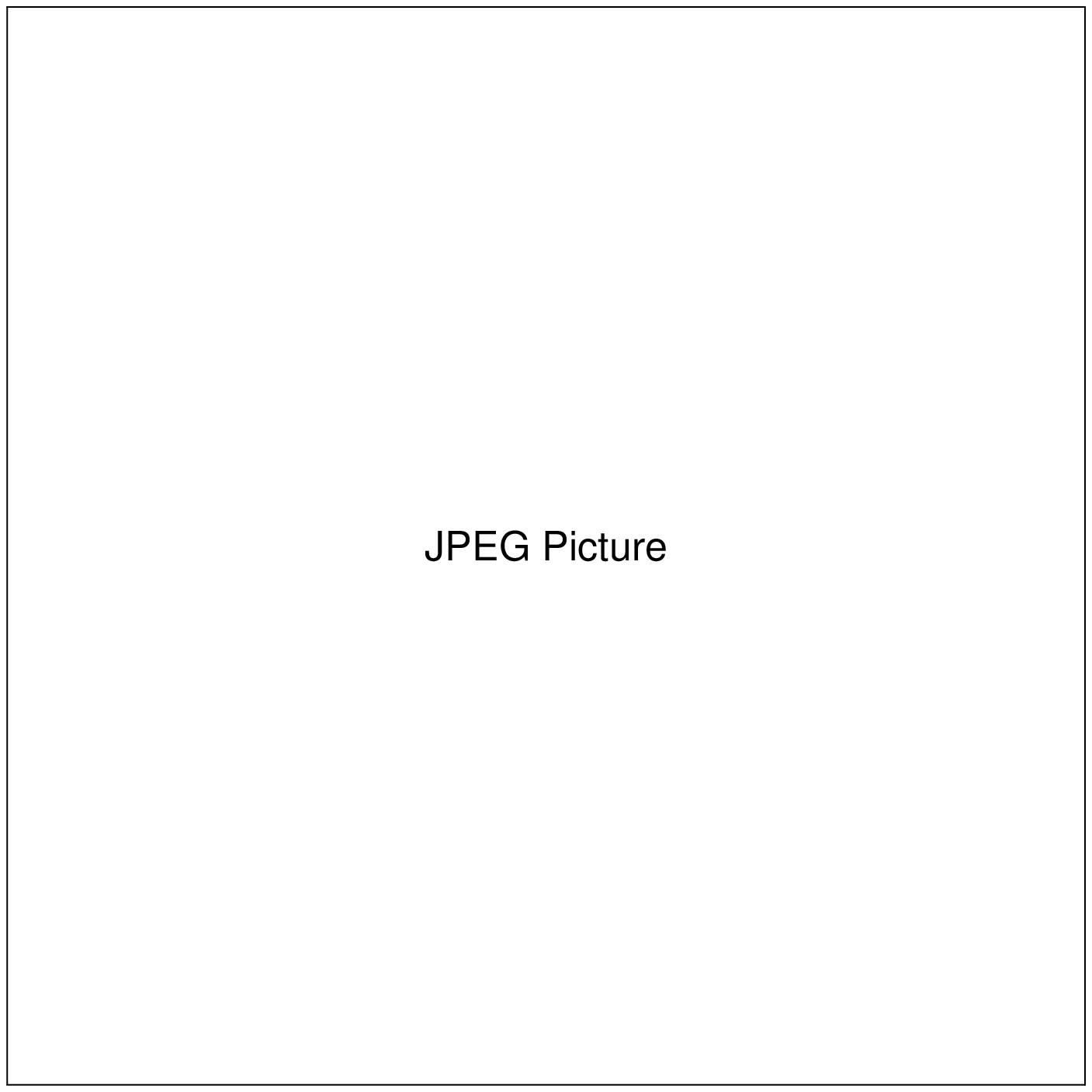}

\plotone{m32pm1_holder.eps}
\caption{{\it V,} {\it I,} and {\it U} deconvolved images of M31
(top, bottom, and top panel on next page).
All panels are shown with an arbitrary logarithmic stretch.
North is $55.7^\circ$ to the left of the top of the pictures.
The region shown is the central $9\asec10\times9\asec10$ (200 PC1 pixels)
centered on P2; however, the {\it V} and {\it I} images are $2\times2$
subsampled.  The {\it U} image is photon-noise dominated, but
the {\it V} and {\it I} images are SBF dominated.}
\label{fig:m31_pic}
\end{figure}

\begin{figure}[bhtp]
\epsscale{0.5}
\plotone{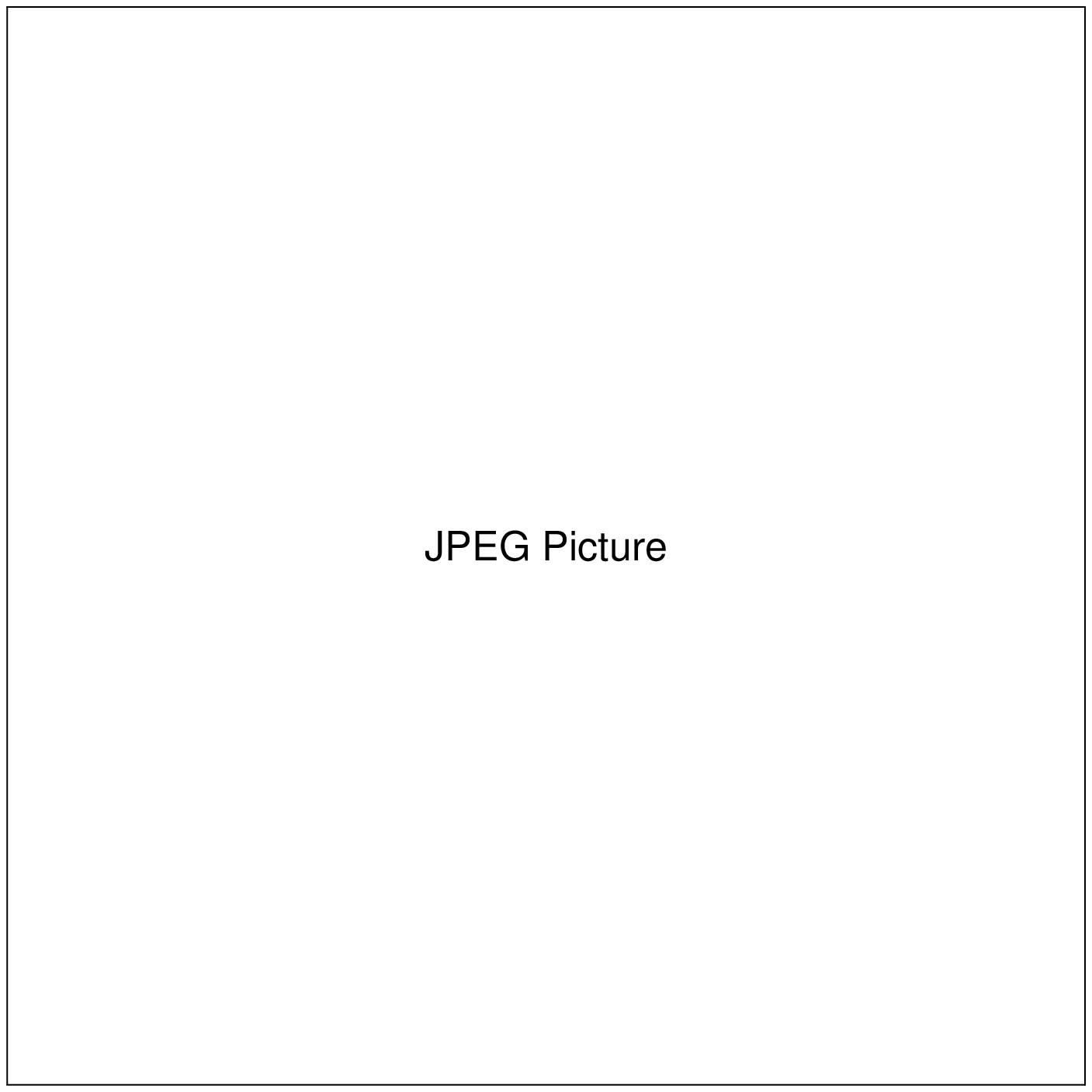}
\caption{A ``true-color'' estimate of the M31 nucleus generated
by using F300W for blue, F555W for green, and F814W for red.  The
intensity stretch is logarithmic.  The region shown is the central
$11\asec65\times11\asec65$ (256 PC1 pixels) centered on P2.
North is $55.7^\circ$ to the left of the top of the picture.}
\label{fig:m31_colpic}
\end{figure}

The F555W, F814W, and F300W images of the M31 nucleus
after deconvolution (with 40 iterations) are shown in
Figures \ref{fig:m31_pic} and \ref{fig:m31_colpic}.
\begin{figure}[htbp]
\plotone{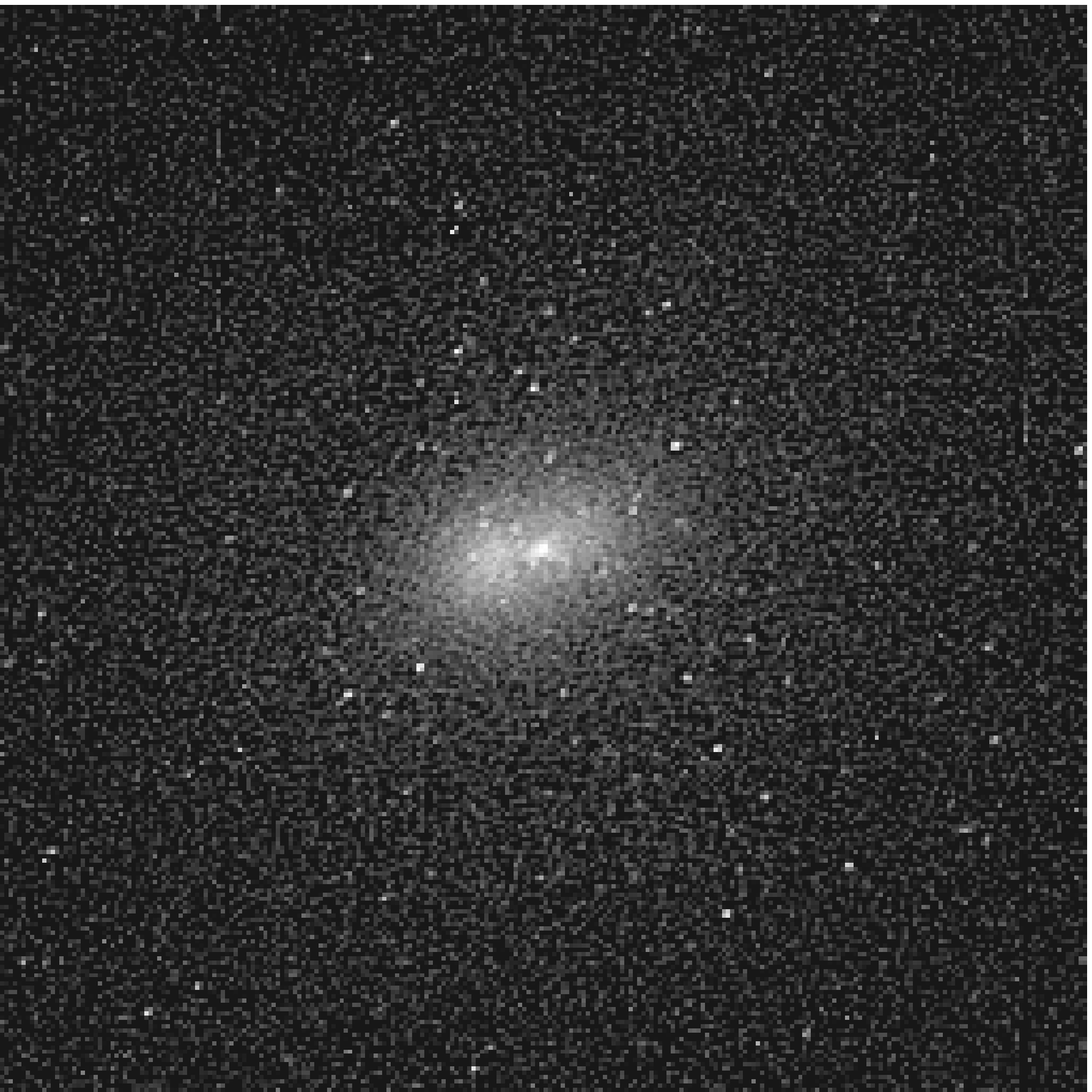}
\end{figure}
Note that the F555W and F814W images have been constructed
from dithered image sets and thus have double the normal
PC pixel scale, while the F300W image is undithered.
The images show the same gross morphology
that was already evident in the WFPC-1 images
(\markcite{l93}Lauer et al.\ 1993);
thus we will not repeat the extensive set of decomposition
experiments that were tried by \markcite{l93}Lauer et al.\ (1993)
and \markcite{k95}King et al.\ (1995).
We do, however, present the V-band brightness profile as measured
with the portion of the image dominated
by P1 excluded from the fit in Table \ref{tab:m31surf}.
One obvious consequence of the improved dynamical range of the new
data is the detection of strong surface brightness
fluctuations in the {\it V} and {\it I} images,
visible as strong mottling (Figure \ref{fig:m31_pic}).
\markcite{ajhar}Ajhar et al.\ (1997)
have measured surface brightness fluctuation
(SBF) amplitudes for the images presented here.
In F555W, the typical SBF ``star'' has $\overline{m}_V=25.8.$\@
At the central surface brightness in P1 of 13.4,
there are thus $\sim190$ SBF stars in each pixel,
giving a relative noise level of 7\%.
In F814W, the surface brightness of P1 is 12.0,
which for SBF $\overline{m}_I=23.4$ implies a noise level of 12\%.
Clearly at {\it HST} resolution, we are running out of the
very stars that trace the structures that we are trying to understand;
the dominant error is not photon noise
but the finite number of stars that make up M31 itself.
SBF is less apparent to the eye at F300W,
but this image clearly contains a number of bright stellar
sources surrounding the nucleus.
The pattern of these sources matches that seen in the
\markcite{k92}King et al.\ (1992)
FOC image taken at F175W; King et al.\ argued that these are PAGB stars.

\begin{figure}[ht]
\epsscale{0.5}
\plotone{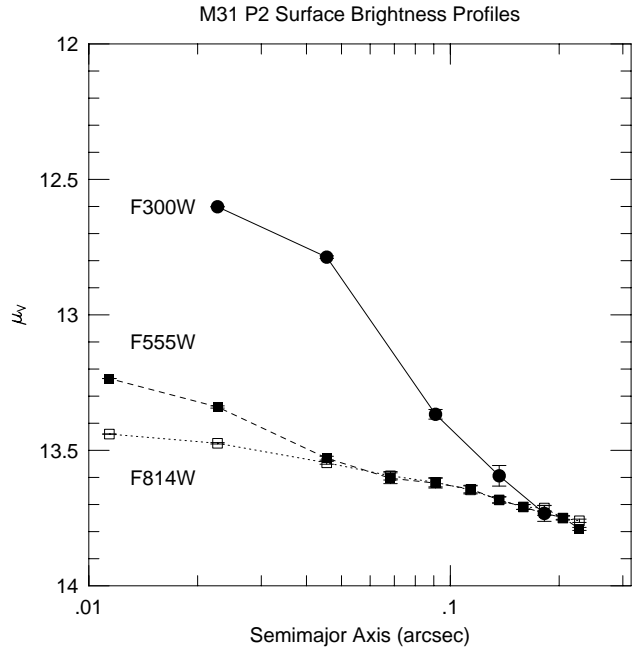}
\caption{Surface brightness profiles of the region in M31
immediately surrounding P2.  The F300W and F814W profiles
are scaled to match the F555W profile at $0\asec2.$}
\label{fig:m31_p2pro}
\end{figure}

Another improvement provided by the WFPC-2 images
is a clearer picture of the structure of P2 and its immediate surroundings.
P2 apparently corresponds to both the center of the galaxy and the
massive black hole inferred to exist within the nucleus (see below).
The F300W image is particularly interesting,
as the dominance of the two peaks
has reversed from that seen in the redder colors;
at 3000 \AA\ a compact blue source is strongly evident,
centered on P2 (Figure \ref{fig:m31_pic}).
\markcite{k95}King et al.\ (1995) already showed that P2 was much
brighter than P1 in the far-UV (at 1750~\AA);
we now see that the cross-over must occur between F300W and F555W.
In the WFPC-2 F555W image, the compact source
at P2 is still evident, if greatly reduced in amplitude.
Normalized surface brightness distributions
centered on P2 are shown in Figure \ref{fig:m31_p2pro}
for the three colors presented here.
The strength of the compact source in F300W is clearly evident;
even at F555W the source causes the profile to
rise above a shallow power-law cusp at $r<0\asec05.$\@
Only at F814W is there no evidence of a source superimposed on the cusp.
In passing, we note that WFPC-1
showed only that the F555W P2 profile consisted of a shallow cusp;
the limited dynamic range of the aberrated observations made it exceedingly
difficult to detect the faint compact central source
against a bright background.

\begin{figure}[ht]
\plotone{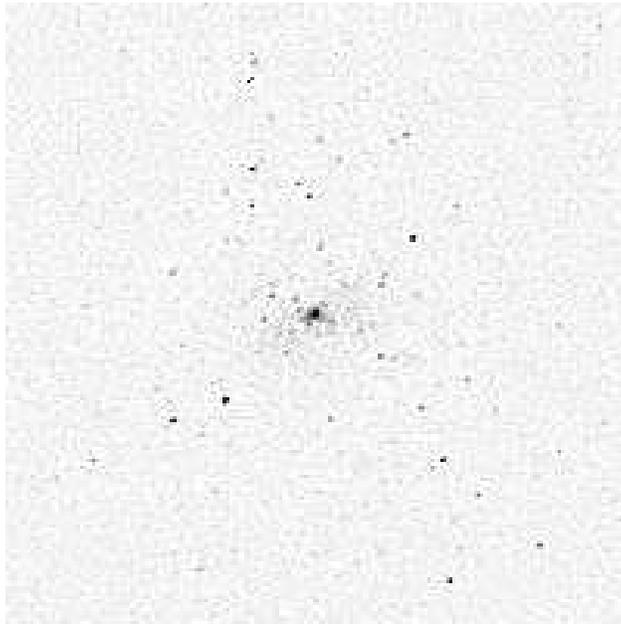}
\caption{A residual image showing the blue source at P2,
creating by subtracting the deconvolved F555W image from the
deconvolved F300W image.
Note that the emission associated with P2 is extended.
The region shown is the central $9\asec10\times9\asec10$ (200 PC1 pixels)
centered on P2.
North is $55.7^\circ$ to the left of the top of the picture.}
\label{fig:m31_p2u}
\end{figure}

The blue P2 source is compact, but not point-like --- it is slightly resolved.
This is evident in Figure \ref{fig:m31_p2u},
where we show the F555W image subtracted from the F300W image
after it was sampled to match the P2 centroid in the
F300W image and scaled to match the surface brightness
of the outer portions of the nucleus.
The residual blue excess of the P2 source clearly
extends over several pixels, while the
stellar sources noted above are only 1-2 pixels in size.
The F300W profile (Figure \ref{fig:m31_p2pro})
shows that the source has a half-power radius of $0\asec06,$ or 0.2 pc.
We estimate its flux by integrating the excess
blue light within a 4 pixel radius in the difference image.
The apparent luminosity of the source is $m_{300}=18.6\pm0.3.$\@
By comparing the F555W and F814W images, we similarly conclude
that the P2 source has $m_V=18.7\pm0.3.$\@
\markcite{k95}King et al.\ (1995) raised the possibility that the P2
source was nonthermal emission from a low-level AGN
and found it convenient to present their F175W luminosity in flux units.
Using the zero-points given in \markcite{holtz}Holtzman et al.\ (1995),
we find $f_{300}=60\pm20~\mu{\rm Jy}$ and $f_{555}=160\pm50~\mu{\rm Jy},$
where we have adopted $A_V=0.24$ (\markcite{bh}Burstein \& Heiles 1984)
and $A_{300}=0.42$ (based on extrapolation from the O6
extinction table of Holtzman et al.).
These fluxes can be compared to
$f_{175}=7~\mu{\rm Jy}$ observed by King et al.
The physical nature of the P2 source is discussed in $\S\ref{sec:dm31}.$\@

\begin{figure}[ht]
\epsscale{0.5}
\plotone{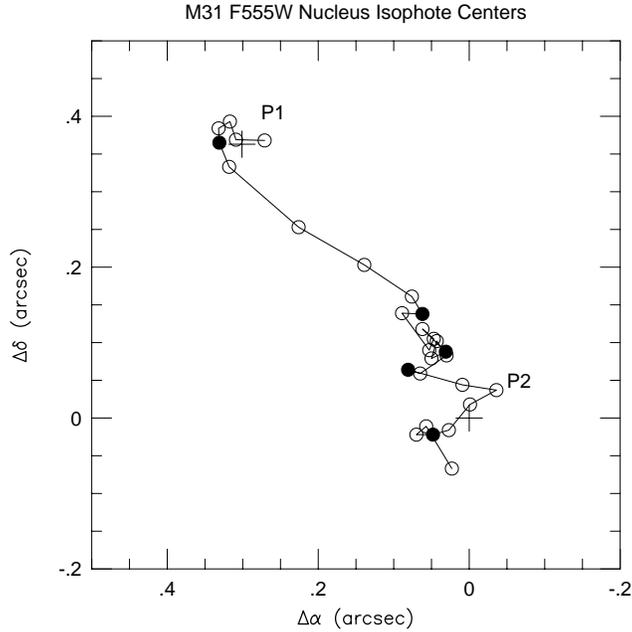}
\caption{The RA and DEC offsets of the isophote centers with
respect to P2 are shown as a function of F555W surface brightness.
The first point plotted near P1 is at $\mu_V=13.1;$ succeeding points step
0.1 mag in surface brightness.
The solid points mark isophotes of $\mu_V=13.5,$ 14.0, 14.5, 15.0, and 15.5.
All isophotes with $\mu_V\geq14.5$ are centered within $0\asec1$ of P2.}
\label{fig:m31_isocen}
\end{figure}

\begin{figure}[ht]
\epsscale{0.5}
\plotone{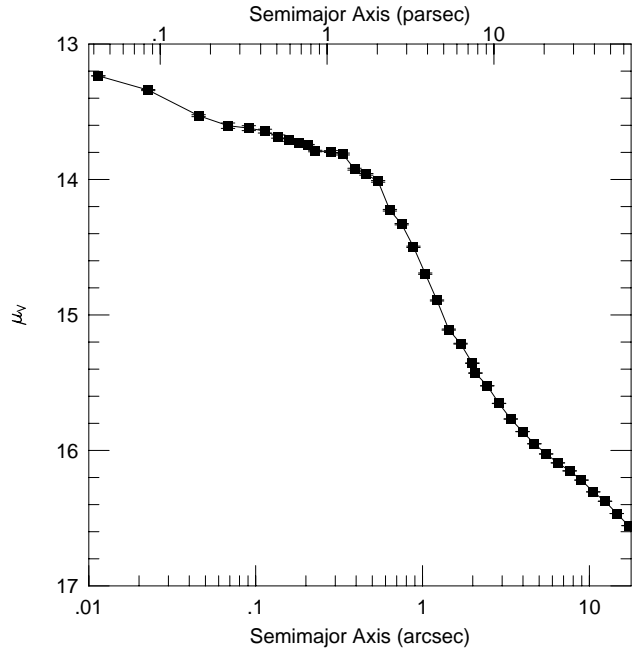}
\caption{The F555W brightness profile of the M31 nucleus.
The portions of the image dominated by P1 have been excluded from
the profile measurement.}
\label{fig:m31_pro}
\end{figure}

The new data confirm the conclusion reached by \markcite{l93}Lauer
et al.\ (1993) that P2 corresponds to the photocenter of M31.
The location of isophote photocenters
as a function of {\it V} surface brightness
is shown in Figure \ref{fig:m31_isocen}.
Centers were calculated by measuring the ``center of mass''
of all pixels within $\pm0.1$ mag of a given isophote level.
Isophotes with $15\leq\mu_V\leq16$ all have centers
within $0\asec1$ of P2.  Indeed, the
centroid of pixels with $\mu_V=15.5\pm0.3$ is only $\approx0\asec05$ from P2.
As can be seen in Figure \ref{fig:m31_pro}, which shows
the central brightness profile of M31 after excluding the area
strongly affected by P1, isophotes of this brightness are $\sim2''$
in radius and thus define the outer portions of the nucleus (that
is the portion of the brightness profile that rises above the bulge profile).
Isophotes at larger radii are affected by dust but also appear
to have centers within $0\asec1$ of P2.
Notably, the offsets seen in these cases are {\it perpendicular} to the
P1-P2 vector (see the lowest surface brightness points
in Figure \ref{fig:m31_isocen});
their projection on the P1-P2 vector is thus still close to P2.

\begin{figure}[ht]
\epsscale{0.5}
\plotone{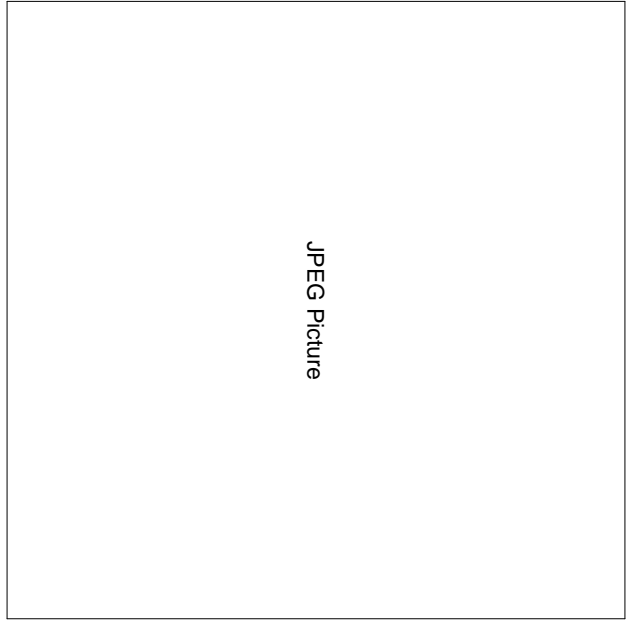}
\caption{The ratio of the M31 F555W and F814W deconvolved images.
The scale is magnitudes, with white corresponding to $V-I=1.1,$
and black to $V-I=1.6.$\@
The region shown is the central $9\asec10\times9\asec10$ (200 PC1 pixels)
centered on P2.
North is $55.7^\circ$ to the left of the top of the picture.
P2 is evident as the white peak at the very center of the image,
surrounded by a region of redder than average color.
The noisy appearance of the image is due to the incipient
resolution of M31 into stars. Tick marks point to the centroids
of P1 and P2}
\label{fig:m31_colrat}
\end{figure}

Color variations appear to show that
the center of the nucleus (the region within
$1''$ of P2) has a slightly different stellar population than
the outer portions of the nucleus and surrounding bulge.
Figure \ref{fig:m31_colrat} shows the ratio of the F555W and F814W images.
P1 is clearly redder in $V-I$ than the surrounding
galaxy, but the region of red excess also extends into the
region surrounding P2 on the opposite side from P1.
The central $0\asec4\times0\asec4$ region of P1 has $V-I=1.41,$
while the color in a similar-sized aperture displaced
$0\asec4$ from P2 in the ``anti-P1'' direction has $V-I=1.38.$\@
The bulge and outer portion of the nucleus are bluer still
with $V-I=1.34$ for $1''<r<5''.$\@
On the other hand, P1 and its anti-region are {\it bluer} than the bulge in 
$F300-V$, with P1 having $F300-V=2.29,$ the anti-P1 part of P2 having
$F300-V=2.18,$ and the isophotes with $1''<r<5''$ having $F300-V=2.39.$\@
In passing, we note that this result is in apparent contradiction
to the results of \markcite{k95}King et al.\ (1995), who found
{\it redder}-than-bulge colors in $F175W-F555W$ from pre-COSTAR FOC
and WFPC-1 data in the region outside $r\sim1''.$\@
For comparison, P2 itself has $V-I=1.28,$ and $F300-V=1.29.$\@
For convenience, 
these color measurements are summarized in Table \ref{tab:m31nuc}.

That the center of the nucleus is redder than the bulge
in $V-I$ yet bluer in $F300-V$ may be reasonable;
such combinations can in principle be created
by simultaneous changes to both stellar population age and metallicity.
However, an indisputable conclusion from the color difference
is that the central nucleus contains
stars with an origin different from the surrounding bulge.
This may support the \markcite{trem}Tremaine (1995)
eccentric disk model of the doubly-peaked nucleus;
the disk would consist of material that has fallen into
the nucleus after the central structure of M31 was already organized.
Additional support for the Tremaine disk may come from the morphology
of the red-excess region evident in Figure \ref{fig:m31_colrat}, given that
it clearly extends to both sides of P2;
P1 would simply be a local density enhancement in a population
with a more extended spatial distribution.
The intermediate color of the anti-P1 part of P2, that is a color between
that of P1 and the bulge, may reflect underlying dilution
of the disk by the bulge.
It is also noteworthy that the intensity minimum between
P1 and P2 that defines the double-peaked appearance of the
nucleus has $V-I=1.34\pm0.02,$ a color bluer than that of
P1 but which matches the bulge.

\subsection{M32\label{sec:om32}}
 
\begin{figure}[htbp]
\epsscale{0.5}
\plotone{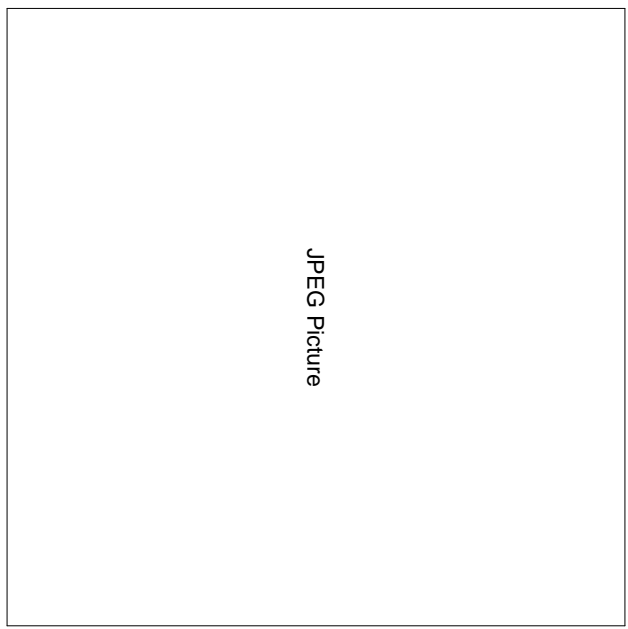}
\caption{The F814W deconvolved image of M32.
North is $120.0^\circ$ to the right of the top of the picture.
The region shown is the central $23\asec3\times23\asec3$ (512 PC1 pixels).
The stretch is logarithmic.}
\label{fig:m32_pic}
\end{figure}

\begin{figure}[bhtp]
\epsscale{0.5}
\plotone{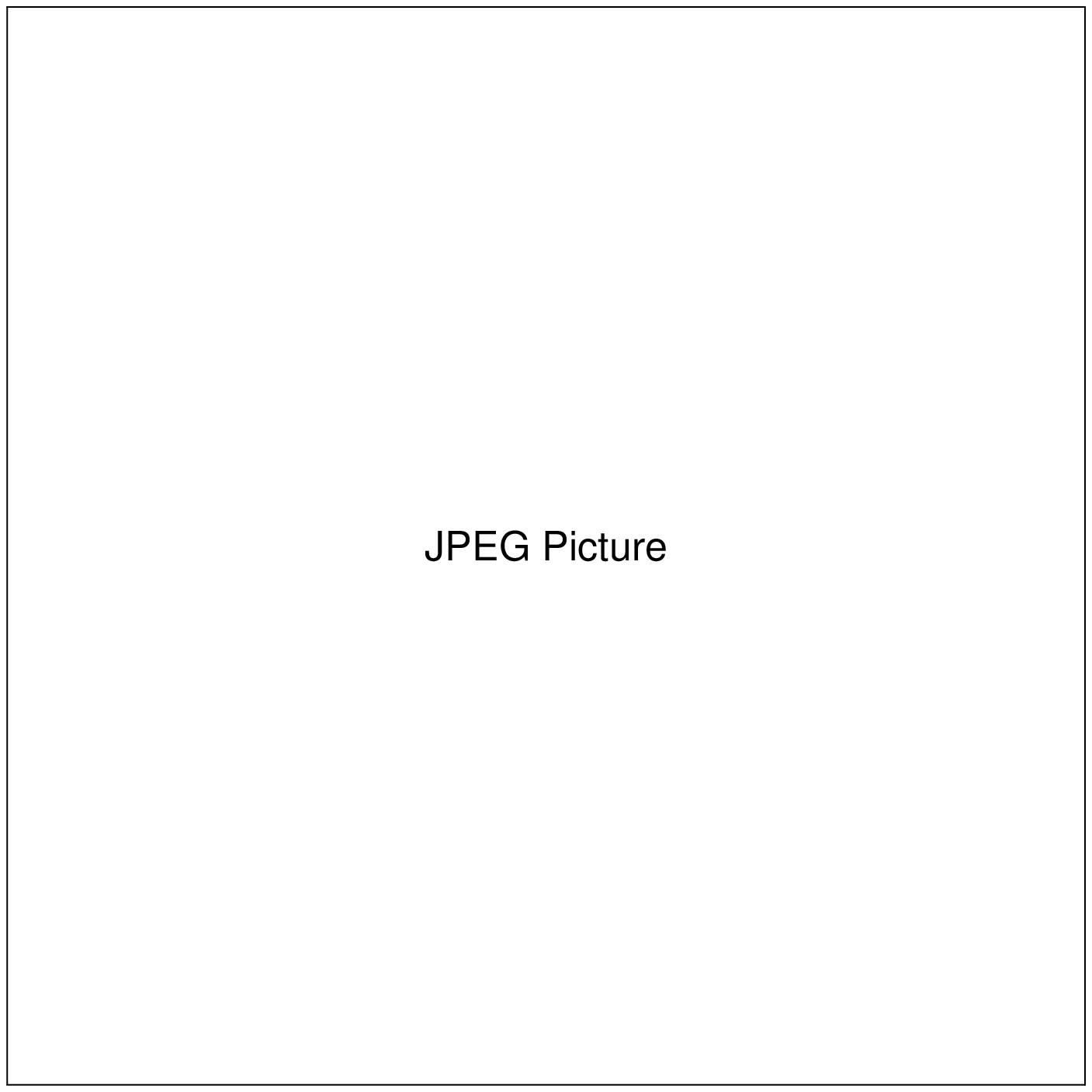}
\caption{A ``true-color'' image of M32 formed from the F555W and F814W
deconvolved images.
North is $120.0^\circ$ to the right of the top of the picture.
The region shown is the central $23\asec3\times23\asec3$ (512 PC1 pixels).
The stretch is logarithmic.}
\label{fig:m32_colpic}
\end{figure}

The F814W image of the M32 nucleus after deconvolution
(with 40 iterations) is shown in Figure \ref{fig:m32_pic};
a color image is shown in Figure \ref{fig:m32_colpic}.
Surface photometry profiles are presented in Figure \ref{fig:m32_pro}
and Table \ref{tab:m32surf}.
The surface photometry was measured with the algorithm of \markcite{kent}Kent
(1983), as modified by \markcite{l85}Lauer (1985)
for high-resolution applications.
The algorithm uses sinc-interpolation and Fourier harmonics
to measure the brightness distribution along isophotes
at every integer-pixel step in semimajor axis length.

The most impressive feature of M32 is its incipient
resolution into stars, as was also seen in M31.
The surface brightness fluctuations in F814W are especially strong;
as in M31, they dominate the profile error budget.
\markcite{ajhar}Ajhar et al.\ (1997) have measured
SBF amplitudes for the images presented here.
In F555W the typical SBF ``star'' has $\overline{m}_V=25.4.$\@
With a central surface brightness of 10.96, or 17.67 per PC pixel,
there are $\sim1,200$ SBF stars in the central pixel,
giving a relative noise level of 2.8\%.
In F814W, $\mu_0=9.79$ and $\overline{m}_I=23.0,$
implying a noise level of 5.0\%.
These errors are close to the deconvolution uncertainties in the central pixel.

\begin{figure}[htbp]
\epsscale{0.75}
\plotone{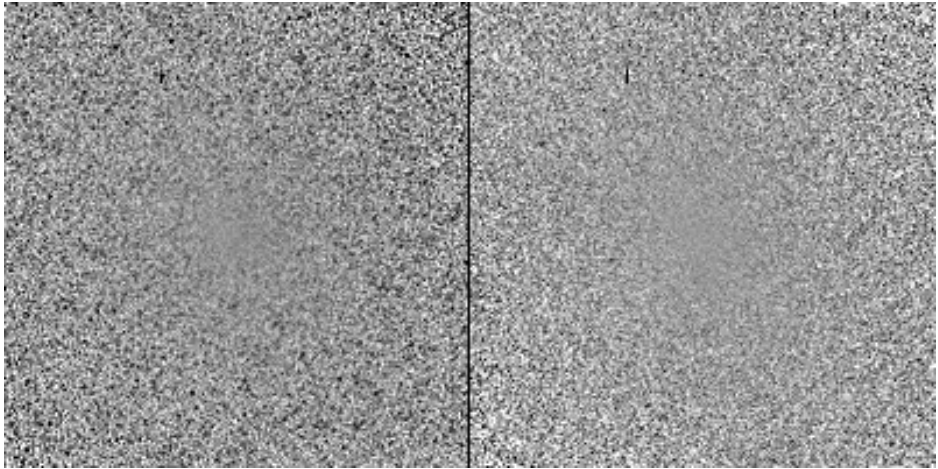}
\caption{The left panel shows the color-ratio image of
M32 formed by dividing the deconvolved F814W image into the F555W image.
The right panel show a residual map formed by subtracting a model reconstructed
from the M32 F555W surface photometry profiles from the original image.
North is $120.0^\circ$ to the right of top in both panels.
The region shown is the central $9\asec10\times9\asec10$ (200 PC1 pixels).
The stretch is $\pm1$ mag is surface brightness about the mean residual.}
\label{fig:m32_colrat}
\end{figure}

\begin{figure}[htbp]
\includegraphics{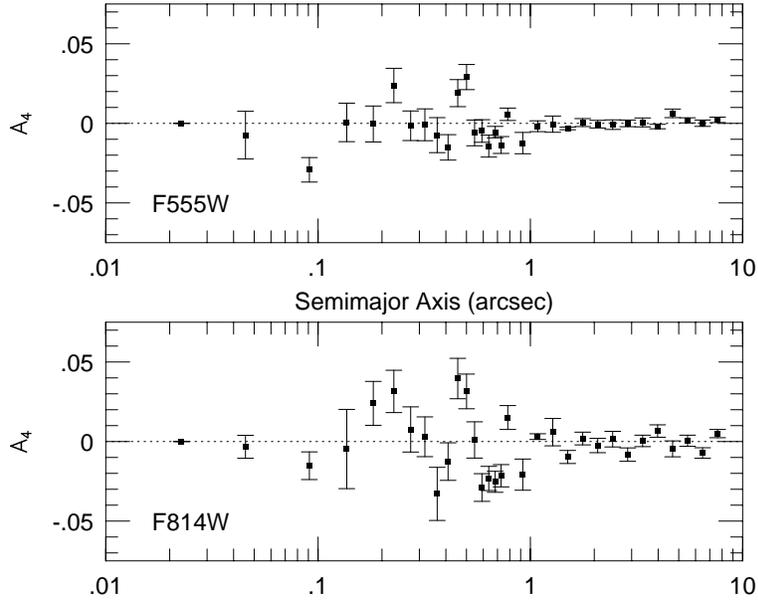}
\vskip 3.0in
\caption{Isophote shape profile of M32 for both the F555W
and F814W deconvolved images. Positive $A_4$ corresponds to ``disky'' isophotes;
negative $A_4$ corresponds to ``boxy'' isophotes.}
\label{fig:m32_a4}
\end{figure}

Aside from the strong SBF pattern,
we find no evidence for dust, disks, or any other structures present
in M32 down to levels of a few percent in local surface brightness.
Figure \ref{fig:m32_colrat} shows a color
ratio image made by dividing the {\it V} image by the {\it I} image,
and the residual map resulting from subtraction of models reconstructed
from the F555W profile.
All such comparisons are flat on average
(although the color-ratio image shows dark ``pits'' due to
the red color of barely resolved bright stars).
The isophote shape profiles additionally show no evidence
for any structures not accounted for by elliptical isophotes.
Figure \ref{fig:m32_a4} shows the $A_4$ isophote shape parameters
\begin{figure}[bhtp]
\epsscale{0.6}
\plotone{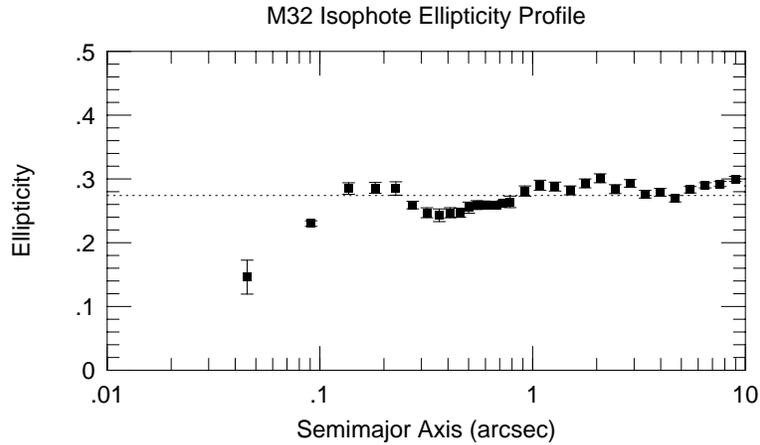}
\caption{Isophote ellipticity profile of the deconvolved F555W image of M32.
The dashed line indicates the average ellipticity for $0\asec1<r<10''.$}
\label{fig:m32_ellip}
\end{figure}
for both F555W and F814W, which give the relative fourth harmonic
deviations of the isophotes from perfect ellipses.
Both profiles are noisy due to strong surface brightness
fluctuations but show no evidence for the isophotes
to be significantly ``boxy'' or ``disky'' (which would
give significantly negative or positive $A_4,$ respectively).
This verifies the conclusions of \markcite{lugger}Lugger et al.\ (1992)
and \markcite{l92}Lauer et al.\ (1992) and stands in contrast
with the claim by \markcite{michard}Michard \& Nieto (1991)
that the inner isophotes of M32 were disky, having positive $A_4.$\@
Michard \& Nieto specifically argued that a disk was 
visible in their reddest images and thus might be comprised of AGB stars.
The F814W profile, while noisier than F555W,
shows no greater evidence for a disk.
It is also noteworthy that the ellipticity of M32 remains
constant throughout the inner portion of the galaxy,
having $\epsilon\equiv1-b/a=0.274\pm0.003$
for $r>0\asec1$ (Figure \ref{fig:m32_ellip}).
\markcite{l92}Lauer et al.\ (1992)
argued that the isophote ellipticity in M32
stayed constant into the resolution limit,
based on deconvolution simulations with WFPC-1;
the constant ellipticity is shown directly here with WFPC-2 data.

\begin{figure}[htbp]
\epsscale{0.6}
\plotone{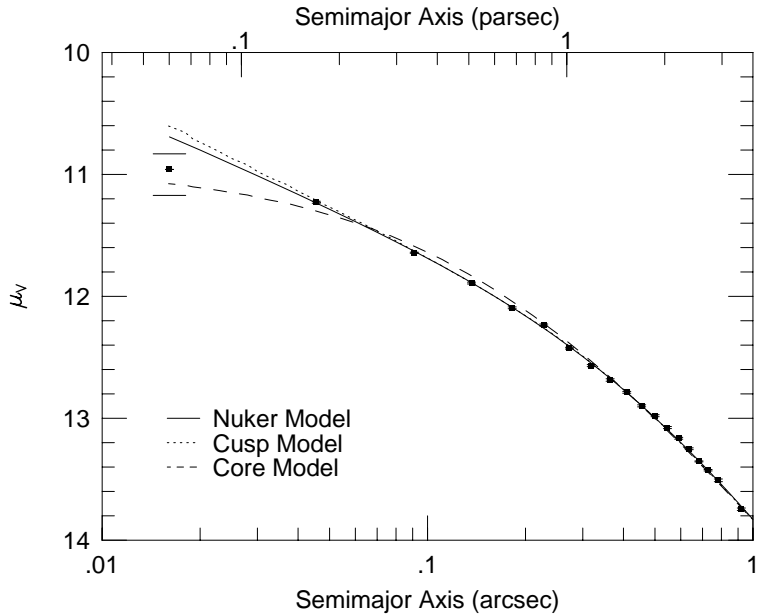}
\caption{The M32 {\it V} surface brightness profile and models.
The ``cusp'' and ``core'' models were introduced
by Lauer et al.\ (1992) and are discussed in the text;
the Nuker model is new to this work.
The horizontal lines bracketing the central point of the deconvolved profile
show the central brightnesses of the core and cusp models accounting for
aliasing; the Nuker model corrected for aliasing is not shown, but
fall slightly below the cusp model.
The Nuker model now gives the best fit.
The inset figure shows the mass-density profile implied
by the Nuker model, assuming $M/L_V=2.0$ in solar units.}
\label{fig:m32_mod}
\end{figure}

As noted in $\S\ref{sec:decon},$ the new WFPC-2 observations of M32
are in excellent agreement with the WFPC-1 F555W profile
of \markcite{l92}Lauer et al.\ (1992) for $r>0\asec1$
but rise above the old WFPC-1 profile at smaller radii due to the
improved resolution of WFPC-2.
The new data allow us to resolve some of the ambiguity in the
central structure of M32 due to uncertainties in the WFPC-1 profile.
\markcite{l92}Lauer et al.\ (1992) showed that the central structure of M32
was well bracketed by a ``core'' model of the form:
\begin{equation}
\label{eq:m32core}
I(r)=I_0\left(1+{r\over0\asec166}\right)^{-1.37},
\end{equation}
where in the {\it V}-band
$I_0$ corresponds to $\mu_0=10.99,$ and a ``cusp'' model
of the form:
\begin{equation}
\label{eq:m32cusp}
I(r)=I_0\left({r\over1''}\right)^{-0.53}\left(1+\left({r\over0\asec378}
\right)^2\right)^{-0.375},
\end{equation}
where $I_0$ corresponds to $\mu_0=13.04.$\@
Both of these models are shown in Figure \ref{fig:m32_mod}.
All but the central point of the {\it V} profile fall on the cusp model.
The core model fits less well, rising above
the observed profile for $0\asec08<r<0\asec3$
and falling below the central two points.
Since the deconvolution simulations show that the profile is accurate to
$\sim1\%$ at all but the central point,
the cusp model can be taken as an excellent description
of the M32 profile for $r>0\asec04,$ or 0.3 pc.

Understanding the light distribution of M32 at smaller
radii clearly depends on how well we can use the central point
in the brightness profile as a constraint.
If we take the effective radius of the central pixel as
$2^{-3/2}$ of the pixel width, or $0\asec016,$ then it
is offering information on the central 0.1 pc of M32.
Superficially, the central point of the {\it V} profile
falls below the cusp model by 0.34 mag.
It appears that this deficit is not due to
residual blurring from incomplete deconvolution.
Simulated observations of the cusp model presented in $\S\ref{sec:decon}$
show that even the central point in the deconvolved F555W profile
can be recovered to $\sim1\%$ with an accurate PSF.
Further, the deconvolved image matches the observed image extremely well
when it is convolved back to the resolution of the latter;
if the deconvolution were incomplete, its central brightness would
be too low to match the observation after this operation.
Fitting the observed image with the PSF convolution of the
deconvolved image plus a scaled PSF to make up for any central
shortfall in the deconvolved image shows that the central pixel
is likely to be underestimated by only $0.01\pm0.01$ mag.
Systematic errors in the PSF are important and will cause corresponding
profile errors larger than this amount.
Using a different PSF that was less well sampled than the
dithered F555W PSF used for the present results
gives a deconvolved profile 0.06 mag brighter in the center.
As noted above, the intrinsic error in the central brightness
itself is 0.03 mag due to the finite number of central stars,
in good agreement with the 0.04 mag of noise measured in
the central $5\times5$ pixel patch by comparing the
image with a model reconstructed from the photometry.
Adding the PSF error and surface brightness error in quadrature
gives the total error in the central pixel of 0.07 mag.

The most serious difficulty with measuring the central brightness is that
at F555W the PC images are undersampled by $\sim2\times,$ and are thus aliased.
Aliasing causes spatial information above the Nyquist sampling frequency to
contribute erroneously to structure at frequencies
{\it below} the sampling frequency.
In simpler terms, it causes irreversible broadening of features.
This in fact appears to be happening at the very center of M32.
Unfortunately, the photocenter of M32 appears to be
significantly displaced from a pixel center.
The CCD-row offset of the photocenter is only 0.15 pixels,
but the CCD-column offset is 0.46, causing significant
broadening of the M32 nucleus in this direction.
To understand the quantitative effect of aliasing,
we constructed an image of the M32 cusp model faithful to
the orientation and centroid of the observation.
The model image was constructed with $3\times$ subsampling of the PC pixels,
with additional subsampling of the central portion of the cusp itself.
When a profile was measured from the synthetic image positioned and binned
to match the PC sampling, the value of the central pixel was fainter by 0.21 mag
than the input model, leaving the
the central difference between the cusp model and observed profile
at only $0.13\pm0.07$ mag, no longer highly significant.
A similar treatment of the core model
shows that, once aliasing is accounted for (and the core
model is rescaled in brightness to match the observed profile
at $\sim0\asec2$), the central observed point exceeds the
central brightness of the core model by $0.21\pm0.07$ mag.
The aliased central points of both models are shown
in Figure \ref{fig:m32_mod} as horizontal lines bounding
the observed central point.

\begin{figure}[htbp]
\epsscale{0.6}
\plotone{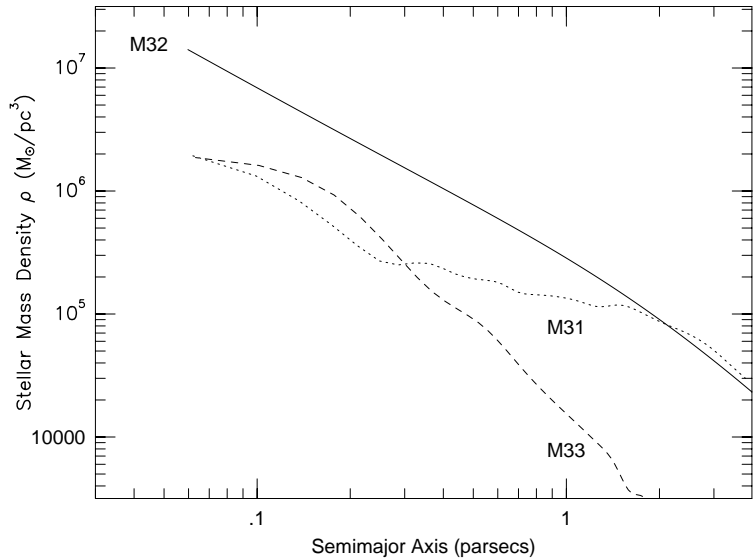}
\caption{Stellar mass density profiles of M31, M32, and M33
based on Abel inversion of the F555W brightness profiles,
assuming stellar population $M/L_V=6.5,\ 2.0,$ and 0.4 in solar units for the
three galaxies, respectively.
For M31 and M33 direct inversions of the deconvolved brightness profiles
are shown; for M32, inversion of the best-fitting Nuker model is shown.}
\label{fig:density}
\end{figure}

The conclusion is that the center of M32 is significantly more
concentrated than the core model.
Despite aliasing corrections, the central
point still falls below the cusp model, although with low significance.
\markcite{l92}Lauer et al.\ (1992) concluded that the asymptotic
slope of the cusp model has $\gamma=0.53\pm0.05.$\@
Taking the corrected central point as accurate implies $\gamma=0.43\pm0.07$
for $<0\asec04.$\@
A Nuker-law fit of the form (\markcite{l95}Lauer et al.\ 1995)
\begin{equation}
\label{eq:nuker}
I(r)=2^{(\beta-\gamma)/\alpha}I_b\left({r_b\over r}\right)^{\gamma}
\left[1+\left({r\over r_b}\right)^\alpha\right]^{(\gamma-\beta)/\alpha},
\end{equation}
to the $V$-profile for $r<1''$ gives $\alpha=1.39\pm0.82,$
$\beta=1.47\pm0.16,$ $\gamma=0.46\pm0.14,$ $I_b$
corresponding to $\mu_b=12.91\pm0.31,$
and $r_b=0\asec47\pm0\asec15;$ this fit is shown in Figure \ref{fig:m32_mod};
note that this model is essentially identical to the cusp model
at all but the central pixel, where it predicts slightly less luminosity.
In passing, we note that the large errors in the fit parameters
include their high covariance; 
the overall form of the model is tightly constrained over the domain fitted
(\markcite{pii}Byun et al.\ 1996).
Taking the Nuker model as the best representation of the profile implies
that the central mass density in stars (Figure \ref{fig:density})is at least
$10^7M_\odot~{\rm pc}^{-3}$ at the {\it HST} resolution limit
(assuming $M/L_V=2.0$ in solar units, $A_V=0.24,$ and 770 kpc distance).
Clearly M32 remains ``cuspy'' at the resolution limit of {\it HST.}
Answering the question of whether we are seeing the cusp
start to flatten, however, 
should be addressed by obtaining new images with sub-pixel dithering.

\begin{figure}[htbp]
\epsscale{0.6}
\plotone{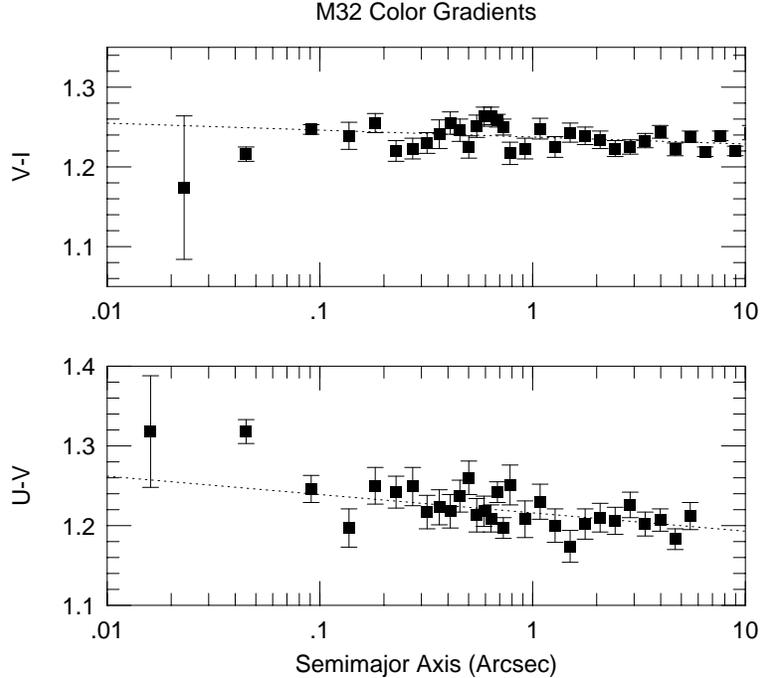}
\caption{$V-I$ and $U-V$ profiles of M32.  The dashed lines
show the mean color gradients for $0\asec1<r<10''.$}
\label{fig:m32_colgrad}
\end{figure}

At larger radii, the WFPC-2 observations 
place excellent constraints on color gradients 
that were not possible with WFPC-1.
As noted in $\S\ref{sec:decon},$ even with WFPC-2,
the effects of the PSF still extend to surprisingly large
radii, and accurate color gradients require deconvolution
and a PSF accurate on large spatial scales.
In the case of M32, it appears that the lack of color gradients
at radii observable from the ground
(for $U-R,$ \markcite{michard}Michard \& Nieto 1991; and $B-R,$
\markcite{lugger}Lugger et al.\ 1992) continues into the center
at least in $V-I.$\@
Figure \ref{fig:m32_colgrad} shows the $V-I$ color profile.
A fit to the $V-I$ trace for $0\asec1<r<10''$
gives $V-I=(1.237\pm0.002)-(0.009\pm0.005)\log(r/1''),$
which is essentially flat.
The color for $r<0\asec1$ becomes tantalizingly bluer.
However, the intrinsic color error
of the central point is 0.06 mag from SBF alone,
and 0.10 mag when PSF errors are included
--- the apparently bluer color is not significant.
As a check on this result, we obtained F336W PC exposures of M32
from the STScI archive and processed them as we did the present observations.
The resulting $U-V$ color gradient is also shown in
Figure \ref{fig:m32_colgrad};
the fitted gradient is $U-V=(1.216\pm0.004)-(0.023\pm0.008)\log(r/1'').$\@
Further, \markcite{cole}Cole et al.\ (1998) found
no far-UV (1600~\AA) versus $V$ gradient for $r<0\asec8.$\@
The $U-V$ colors should be a more sensitive test for hot stars at the
center of M32 than $V-I;$
as the $U-V$ color becomes {\it redder} towards the center,
we find no evidence that M32 harbors any population similar
to that at the center of P2.
From the errors, we estimate that any blueing trend in the central pixel of
M32 is no more than 15\% of the $U-V$ blueing seen within $r = 0.2$ pc in M31.
 
\subsection{M33}
 
\begin{figure}[htbp]
\plotone{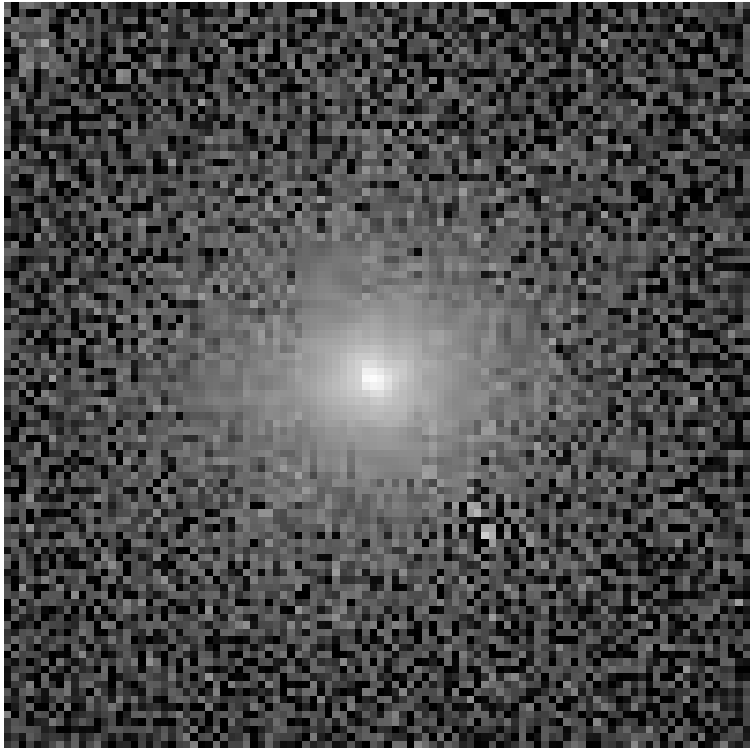}
\caption{The F555W deconvolved image of M33.
North is $113.4^\circ$ to the right of the top of the image.
The region shown is the central $4\asec55\times4\asec55$ (100 PC1 pixels).
The stretch is logarithmic and covers 10 magnitudes in surface brightness.}
\label{fig:m33_pic}
\end{figure}

\begin{figure}[thbp]
\epsscale{0.5}
\plotone{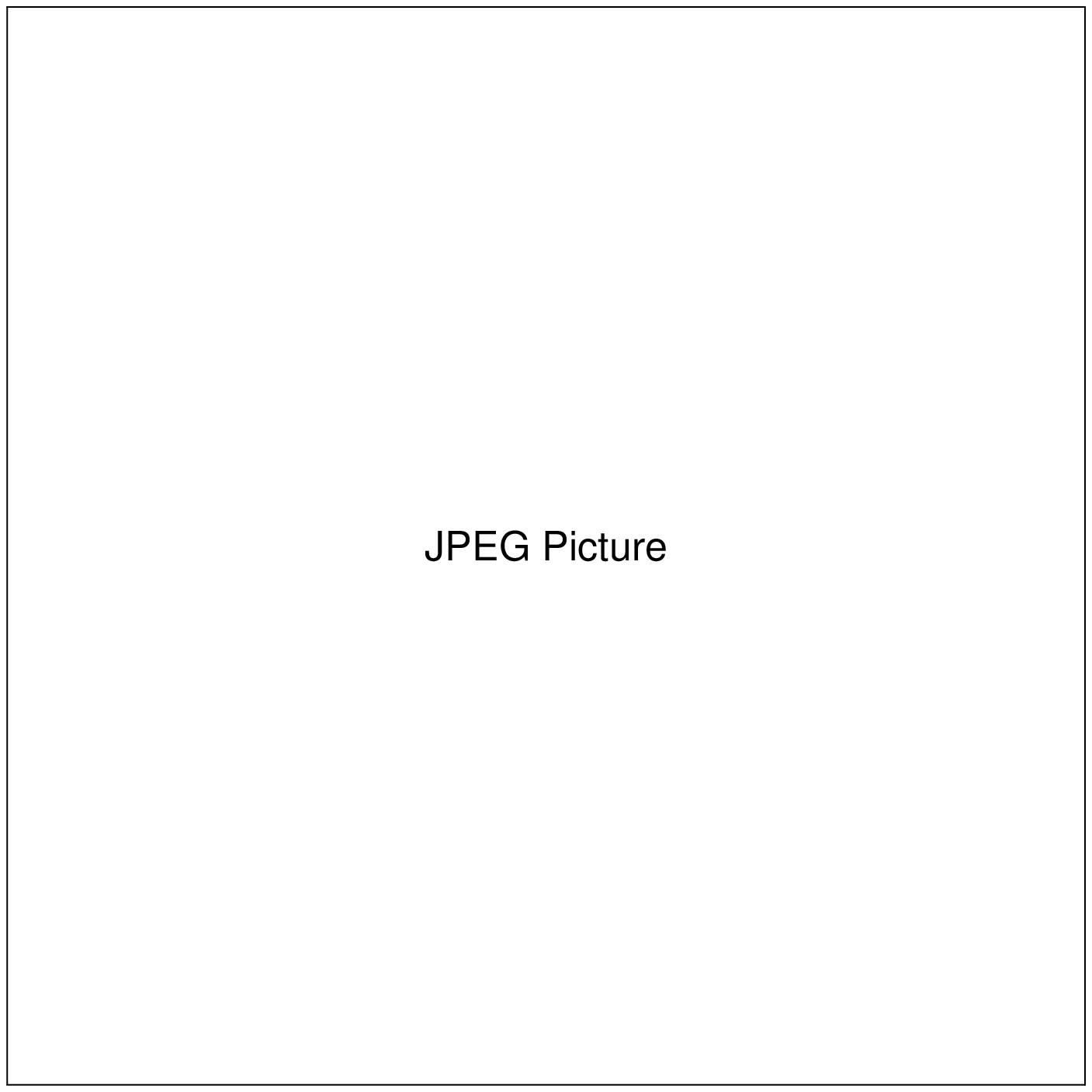}
\figcaption{A ``true-color'' image of M33 formed from the F555W and F814W
deconvolved images. North is $113.4^\circ$ to the right of the top of the image.
The region shown is the central $11\asec7\times11\asec7$ (256 PC1 pixels).
The stretch is logarithmic.}
\label{fig:m33_colpic}
\end{figure}

The F555W image of the M33 nucleus after deconvolution is shown in
Figure \ref{fig:m33_pic}; a color image is shown in Figure \ref{fig:m33_colpic}.
Surface photometry profiles are presented in
Figure \ref{fig:m33_pro} and Table \ref{tab:m33surf}.
Both the images and surface photometry plots show vividly that
M33's nucleus is extremely compact.
While the central F555W surface brightness is brighter than 11 mag/\sq\ arcsec,
the brightness profile falls rapidly as
a power law with slope $\gamma=1.49\pm0.09$ for the first semi-decade in radius
($0\asec08<r<0\asec25$), steepening to $\gamma=1.90\pm0.03$ over the
next semi-decade ($0\asec25<r<0\asec8$).
At $r\sim2''$, the nucleus profile begins to fall below the disk,
having decreased to $\mu_V>18.5,$ or $\sim10^3$ dimmer than the central value.
At $r\sim1'',$ the nucleus may be showing signs of incipient resolution into
stars, but as the disk begins to dominate, it is difficult to associate
any of the point sources at larger radii with the nucleus itself.

We first consider the WFPC-2 constraint on the central luminosity density.
\markcite{km}Kormendy \& McClure (1993) argued that the M33 nuclear core radius
must have $r_c<0\asec10,$ based on seeing-convolved models.
As shown in Figure \ref{fig:m33_pro},
the present observations match the best Kormendy \& McClure {\it B}-band
profile for $r>0\asec3$ (assuming $B-V=0.45$),
but climb to surface brightnesses over an order of magnitude
brighter at small radii.
Near the {\it HST} resolution limit, the M33 profile is still climbing steeply.
If we take the deconvolved profile as accurate excluding the
central point, then the first and second points out give $\gamma=1.44,$
consistent with the slope measured at $0\asec08<r<0\asec25.$\@
The central point {\it a priori} is most likely to be a lower-limit
on the true central surface brightness.
Deconvolution experiments on simulated M33 profiles 
show that we can recover the central intensity
at the few percent level, provided that the PSF is accurate;
however, as with M32, aliasing of the images is important.
The photocenter of M33 in the F555W image is displaced 0.41 pixels
in the CCD row direction and 0.30 pixels in the column direction
from the center of a pixel.

As with M32, we constructed models with steep central
power-law profiles on a subpixel grid, and then binned the models
to match the observations.
For cusps with $\gamma\sim1,$
the aliasing of the M33 image 
depresses the central intensity from its true value by over 0.2 mag.
The procedure for estimating the central brightness of M33 was to
assume that its intrinsic profile continues inward
as a constant power-law for $r<0\asec04$
and construct models for a range of slopes that produced
central decrements bracketing the observations.
The best fitting model gives $\gamma=0.82\pm0.17$
after accounting for aliasing;
the uncertainty reflects an adopted error in the observed
central intensity of 0.07 mag due to PSF uncertainties
and surface brightness fluctuations.
This result suggests that the M33 nucleus profile
is still becoming shallower near the {\it HST} resolution limit,
continuing the slow decrease in logarithmic slope
seen over $0\asec08<r<0\asec8.$\@
Inward extrapolation of the $\gamma=1.49$ slope seen
just outside the center predicts an intensity decrement
between the central and next pixel in the profile of 0.94
magnitudes, compared to the observed decrement of only 0.59 mag.

In contrast, \markcite{km}Kormendy \& McClure (1993)
fitted their profiles to a form that included an ``analytic'' core,
that is, one that requires $\gamma\rightarrow0$ smoothly as $r\rightarrow0.$\@
In this spirit, we also modeled the central portion
of the profile with the form:
\begin{equation}
\label{eq:m33}
I(r)=I_0\left(1+\left({r\over a}\right)^2\right)^{-1.49/2}.
\end{equation}
Fits to the {\it V-}band image give $a=0.6\pm0.1$ PC pixels.
Defining $r_c$ as the half-power point implies $r_c=1.24a,$
thus $r_c=0\asec034\pm0\asec006,$ or $0.13\pm0.02~{\rm pc}$
for a distance to M33 of 785 kpc.
This form is no more favored than the constant $\gamma=0.82$
cusp; thus this core size should be regarded only as an upper limit.
We note that \markcite{mass}Massey et al.\ (1996)
claimed $r_c=0\asec08$ from the present images obtained from
the {\it STScI} archive; however, they neglected to account
for the effects of the PSF or aliasing.
\begin{figure}[htbp]
\epsscale{0.6}
\plotone{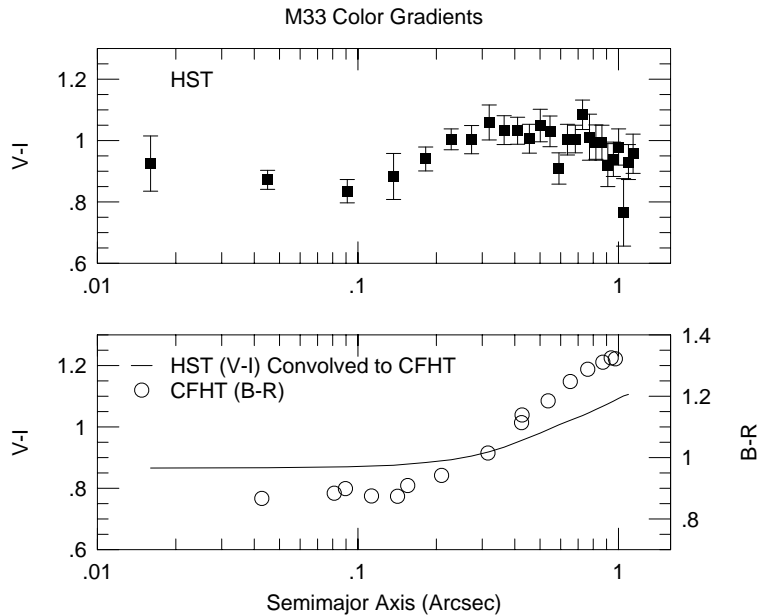}
\caption{Central color gradients in M33.
The upper panel shows the difference between the {\it V} and
{\it I} profiles as a function of radius (uncorrected for reddening).
The bottom panel compares the present $V-I$ profile degraded to CFHT seeing
(solid line) to the $B-R$ profile of Kormendy \& McClure (1993) (open circles).}
\label{fig:m33_colgrad}
\end{figure}

Abel inversion of the M33 brightness profiles implies that
its nucleus reaches central mass densities of
at least $2\times10^6M_\odot~{\rm pc}^{-3}$ (Figure \ref{fig:density}).
While the peak luminosity density of M33 exceeds that of M32,
\markcite{km}Kormendy \& McClure (1993) conclude that M33 has central
$M/L_V\leq0.4,$ a value $\sim5\times$ less than that
of the stellar population of M32.
M32 thus reaches much higher mass-densities on the scales probed by {\it HST.}

The structural analysis was performed on the {\it V}-band
image, as it has higher spatial resolution than the {\it I-}band image.
Comparison of the {\it V} and {\it I} profiles reveals
a strong central color gradient,
which appears to be consistent with the $B-R$ color
gradient observed by \markcite{km}Kormendy \& McClure (1993) at CFHT resolution.
Figure \ref{fig:m33_colgrad} shows the $V-I$ color profile obtained from the
two deconvolved profiles.  The color appears to be constant
at $V-I=1.00\pm0.01$ for $r>0\asec3$ but becomes bluer,
with $V-I=0.85\pm0.02,$ closer to the center.
The low surface brightness of the nucleus at $r\sim1''$ yields large
errors with the fine-resolution brightness profiles used to
generate the color profiles;
using larger bins reduces the noise but still shows that
$V-I$ is essentially constant for $r>0\asec3.$\@
In passing, we note that even $V-I=1.0$ is significantly
bluer than the typical $V-I\sim1.3$ for the central regions
of elliptical galaxies (\markcite{l98}Lauer et al.\ 1998),
consistent with the well known blue color of the M33 nucleus
(\markcite{vdb}van den Bergh 1991).
These colors are as observed, without reddening corrections.
For $A_B=0.18$ (\markcite{bh}Burstein \& Heiles 1984), $\Delta(V-I)=0.05,$
giving the central intrinsic $V-I=0.80$ and $V-I=0.95$ at larger radii.

Kormendy \& McClure find $\Delta(B-R)\approx0.4$ over $0\asec2<r<1\asec0.$\@
This gradient is larger in amplitude and
occurs at larger radii than the $V-I$ gradient,
but their data were not corrected for seeing.
When the present data are degraded to the best CFHT seeing
of $0\asec44,$ the new $V-I$ profile appears constant for $r<0\asec2$
\begin{figure}[htbp]
\epsscale{0.6}
\plotone{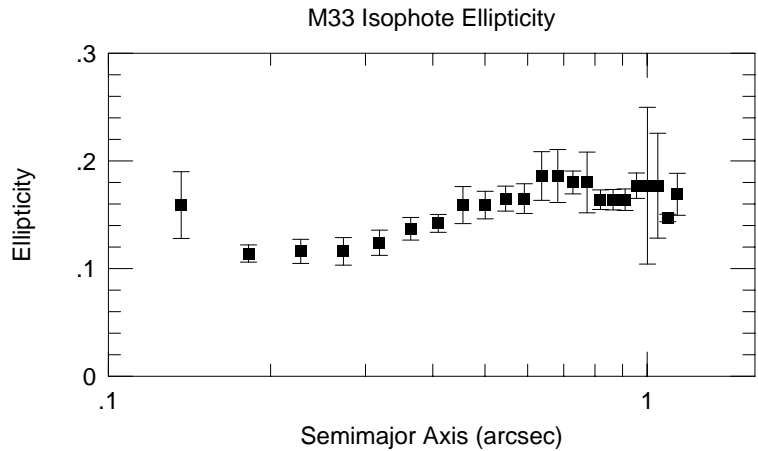}
\caption{Isophote ellipticity profile of the deconvolved F555W image of M33.}
\label{fig:m33_ellip}
\end{figure}
but shows an essentially linear gradient for $0\asec2<r<1\asec0,$
in good agreement with the general shape
of the Kormendy \& McClure $B-R$ profile (Figure \ref{fig:m33_colgrad}).
The present results thus confirm the Kormendy \&
McClure result that the nucleus becomes bluer towards the center.
The larger $\Delta(B-R)$ compared to $\Delta(V-I)$ 
is consistent with the central population containing 
hotter stars, as discussed in the next section.

The surface photometry of M33
also includes information on the shape of the nucleus.
Inspection of the images shows that the nucleus
appears to be slightly flattened, which is confirmed by the
isophote ellipticities plotted in Figure \ref{fig:m33_ellip}.
The average isophote shape over the central arcsecond
has $\epsilon=0.16\pm0.01,$
with the nucleus becoming somewhat rounder towards its center.
The position angles of the isophotes also appear to be constant,
with average $PA=17^\circ\pm2^\circ.$\@
As discussed later, the relaxation time in the M33 nucleus is extremely short;
the ellipticity of the nucleus is probably more likely due to
rotation rather than an anisotropic velocity ellipsoid.
\markcite{km}Kormendy \& McClure (1993) find a small velocity gradient
of $6\pm2$ km s$^{-1}$ arcsec$^{-1}$ (with slit $PA=23^\circ$), 
suggestive of rotation, which can
be compared to the central velocity dispersion of $21\pm3$ km s$^{-1}$.
The slightly flattened shape of the nucleus appears to be consistent
with this modest amount of rotation.

\section{Discussion}

\subsection{M31\label{sec:dm31}}

A key advance provided by the WFPC-2 images of M31 is the detailed look
at the color and morphology of P2.
As noted above, \markcite{k95}King et al.\ (1995) first observed the far-UV
dominance of P2; the present data show that P2 dominates at
3000~\AA\ as well and is 
much bluer in $V-I$ than P1, the rest of nucleus, and the surrounding bulge.
As noted by \markcite{l93}Lauer et al.\ (1993) and confirmed here, 
P2 closely corresponds
to both the dynamical and photocenter of M31, and thus is likely
to coincide with the $3\times10^7M_\odot$ black hole inferred to exist
at the center of M31 \markcite{kr}(Kormendy \& Richstone 1995).
King et al.\ were thus tempted to consider that
the UV flux of P2 was optical emission associated with a low-level AGN.
In support of this hypothesis they cited the radio source
observed by \markcite{crane}Crane, Dickel, \& Cowan (1992) and the X-ray source
observed by \markcite{trin}Trinchieri \& Fabbiano (1991), both coincident
with the center and both having flux densities that
bracketed the UV luminosity of P2.
At the same time, King et al.\ realized that this picture
probably required P2 to be a point source in the UV,
an issue that was difficult to test with the aberrated FOC data.

The present images show that P2 actually has a half-light radius of
$\approx0.2~{\rm pc}$ at 3000~\AA\ and is clearly extended in $V$ as well.
This result suggests the more prosaic possibility raised
by King et al.\ that there is tightly bound cluster
of early-type stars associated with P2.
The range of fluxes from 1750~\AA\ to $V$ can potentially be
explained by a population with $T_e\sim10^4$ K,
which corresponds to late B to early A-type spectral class.
This is in good agreement with the FOC F175W and F275W flux
measurements by \markcite{brown}Brown et al.\ (1998),
who estimate $T_e=1.2\times10^4$ K, for P2.
The smooth appearance of the P2 cusp implies that the
stars are individually faint compared to the total luminosity;
a turnoff at $M_V\approx0,$ corresponding to $T_e\sim10^4$ K,
would be suitably fainter than the total P2 blue source magnitude of 
$M_V\approx-5.7.$\@

Although a cluster of warm stars might explain the blue center
of P2, the origin of such a cluster is somewhat problematic.
One obvious possibility is that the cluster consists of recently formed stars.
The presence of gas and dust in the
bulge of M31 would provide raw material for such stars.
However, at 0.2 pc,  the nominal core of the blue population, the
local circular velocity about the $3\times10^7M_\odot$ black hole
is $\sim800$ km s$^{-1}$, and the rotation curve is Keplerian with strong shear;
it is difficult to picture even a disk of dense gas here as a favorable
site for recent star formation (we discuss this issue further
in $\S\ref{sec:conclude}$).
 
It may be possible that stellar collisions play a role
in explaining the blue population.
King et al.\ (1995) briefly considered this picture but
dismissed it without quantitative discussion;
however, the new stellar density profile suggests
that this idea merits closer attention.
In general, the mean time for any star to suffer a collision
in a Maxwellian distribution of identical stars is
\begin{equation}
\label{eq:tcoll}
t_c=\left[16\sqrt{\pi}n\sigma r_\ast^2\left(1+\Theta\right)\right]^{-1},
\end{equation}
where $n$ is the stellar {\it number} density,
$\sigma$ is the 1-d stellar velocity dispersion, $m_\ast$ is the stellar mass, 
$r_\ast$ is the stellar radius,
and $\Theta\equiv Gm_\ast/(2\sigma^2r_\ast) = (V_\ast^{esc}/2 \sigma)^2$ 
is the ``Safronov number'' (\markcite{bt}Binney \& Tremaine 1987).
Abel inversion of the F555W profile centered on P2 implies that,
near the {\it HST} resolution limit, 
the P2 stellar mass density has the form
$\rho\approx 1.3\times10^6(0.1~{\rm pc}/r)^2 M_\odot~{\rm pc}^{-3},$ given the
bulge $M/L_V=6.5$ (\markcite{k88}Kormendy 1988), $A_V=0.24,$ and 770
kpc distance (Figure \ref{fig:density}).
For $\sigma=7.2\times10^2(0.1~{\rm pc}/r)^{1/2}$ km s$^{-1}$ implied by the
nominal black hole
\footnote{For an $r^{-3/2}$ density cusp surrounding a black hole,
$\sigma=\sqrt{\langle V^2/3\rangle}=\sqrt{2GM_\bullet/(5r)},$
where $M_\bullet$ is the black hole mass (\markcite{y80}Young 1980).
A cusp of this type is not strictly correct for P2, but it
is an excellent match to M32.}, 
the mean collision time for solar-type
stars is $t_c\approx6\times10^{10}(r/0.1~{\rm pc})^{5/2} {\rm yr}$
(this calculation is summarized in Table \ref{tab:time}).
For a stellar population with lifetime $10^{10}$ yr,
roughly 16\% of main sequence turnoff stars would collide,
and thus roughly $2 \times$ 16\% of the total stellar luminosity
(see $\S$\ref{sec:dm32}) would be radiated
from blue-straggler-type stars of higher surface temperature.
Within the accuracy of this calculation, this is not far from the 20\% excess
light contributed by the blue population of P2 in the V-band.
 
On the other hand,
the stellar density profile presented in Figure \ref{fig:density}
presumes constant $M/L_V,$ which will not to be valid if P2
has been modified by stellar collisions;
the central density will be overestimated and thus the
collision time underestimated.
The central blue part of P2 is clearly visible in the density profile
as a distinct inflection;
naive inward extrapolation of the outer part of the P2 cusp
yields a more modest $\rho\approx 4\times10^5 M_\odot~{\rm pc}^{-3}$ and
$t_c\approx2\times10^{11}{\rm yr}$ at 0.1 pc.
With this collision time, the contribution of blue light
would be only $\sim 10$\%.
Furthermore, the $10^4$ K effective temperature of the warm component,
corresponding to a mean spectral type earlier than A0,
is hotter than blue stragglers seen in globular clusters,
which are typically late A or F stars.
It is not clear that single collisions of two main-sequence turnoff stars
can yield stars this hot (\markcite{bail}Bailyn 1995).
Hotter, more massive merger products could of course be produced by
collisional runaway caused by the sinking of massive merged stars
to the center.  However, the small fraction of light involved in
the P2 blue excess suggests that no such runaway has occurred.
Moreover, the dynamical relaxation time at 0.1 pc in M31
is $\sim10^{12}$ yr on account of the high velocity dispersion
near the black hole, further evidence against such a runaway.
Finally, there is no guarantee that massive blue stragglers
would actually form from the energetic stellar collisions in M31,
as the star-star impact velocity at 0.1 pc is $\sim$1800 km s$^{-1}$
(assuming a 1-d velocity dispersion of 700 km s$^{-1}$
and an isotropic velocity ellipsoid).
This exceeds $2\times$ the main-sequence
stellar escape velocity of $\sim 600$ km s$^{-1}$,
the point where major mass loss sets in in stellar
collision models (\markcite{ben}Benz \& Hills 1988).

Note, however, that all preceding arguments 
implicitly assume a single hot stellar
population with a more-or-less isotropic velocity ellipsoid.
In $\S$\ref{sec:conclude} we revisit these estimates under the assumption
that a cold disk of stars is present in the nucleus and find that
the collision time in such a system could be shorter and more favorable.
Such a cold disk is a key assumption in the \markcite{trem}Tremaine (1995)
eccentric-disk model of the nucleus.  Evidence in favor
of it comes from both the bluer population at the center of P2 and the red color
of the inner portions of the nucleus.
The former evidence favors young stars and suggests that significant
infall of gas
to the center of M31 may have occurred well after its initial formation.
The latter evidence indicates that the population of the nucleus
differs significantly from the bulge, again supporting
its formation as a separate event.
the color map is much more symmetrical than the intensity distribution.
The color uniformity makes it more difficult to
accept P1 as a separate stellar system grafted onto
the nucleus rather than as a simple rearrangement of the basic
nuclear stellar population.

\subsection{M32\label{sec:dm32}}

The previous WFPC-1 observations of M32 admitted a wide variation in possible
central structure for $r<0\asec1$ (\markcite{l92}Lauer et al.\ 1992).
The new observations show that the surface brightness of
M32 is still rising as a $\gamma\approx0.5$ cusp interior
to $r<0.1~{\rm pc,}$ reaching $\rho>10^7M_\odot~{\rm pc}^{-3}$ in stars alone.
The corresponding lower limit from WFPC-1 was only
$\rho_0\approx4\times10^6M_\odot~{\rm pc}^{-3}.$\@
At the same time, the WFPC-2 images extend to greater sensitivity
and smaller radii the conclusion that the inner structure of M32
is featureless aside from the inward continuation of a strong cusp.
The center of M32 has no detectable disk, dust, boxy isophotes, or isophote
shape changes.
The $V-I$ and $U-V$ color gradients are nearly flat, showing no dramatic
change in the stellar population as one approaches the postulated
central massive black hole.
 
The recent FOS observations of \markcite{vdm97}van der Marel et al.\
(1997) show that the velocity dispersion continues to rise into the
center at {\it HST} resolution,
thus the rapid relaxation implied by the \markcite{l92}Lauer et al.\ (1992)
``core model'' without a black hole can be ruled out.
In general, the relaxation time scale is
\begin{equation}
\label{eq:trelax}
t_r={0.34\sigma^3\over G^2m_\ast\rho\ln(0.4N)},
\end{equation}
where $\rho$ is the stellar mass density and $N$ is
the number of stars in the system (\markcite{bt}Binney \& Tremaine 1987).
The present observations show that the central mass density profile has the form
$\rho(r)\approx7\times10^6\left(r/0.1~{\rm pc}\right)^{-3/2}M_\odot~{\rm pc}^{-3}.$\@
For $\sigma(r)=242\left(r/0.1~{\rm pc}\right)^{-1/2}$ km s$^{-1}$ as implied
by the $3.4\times10^6M_\odot$ black hole mass of
\markcite{vdm98}van der Marel et al.\ (1998),
$t_r\approx3\times10^9$ yr throughout the center of M32
(assuming $N\sim10^6$), too long to have a significant structural effect.

\markcite{l92}Lauer et al.\ (1992) argued that 
a black hole of this magnitude would
greatly increase the collision time scale over that calculated for M32 without
a black hole; however, they also noted that significant collision effects might
still be expected at scales poorly probed by the WFPC-1 observations.
From equation (\ref{eq:tcoll}) and the $\sigma$ and $\rho$ profiles
presented in the previous paragraph, we estimate $t_c\approx2\times10^{10}$ yr
at 0.1pc, increasing outwards roughly linearly with radius
(see Table \ref{tab:time}).
 
This new central collision time scale in M32 may be low enough to
imply detectable changes in the spectral energy distribution if 
the stellar population is old.
We assume that collisions acting
on a base population of turn-off stars of age $t_{to}$ and mass $M_{to}$ are
creating blue stragglers of lifetime $t_S \sim 10^9$ yr and mass $M_S$.
Let the bolometric luminosity
of the blue stragglers be $L_S$ and that of the turn-off stars be $L_{to}$.
If the system is older than the blue-straggler lifetime, then 
creation of blue-stragglers from collisions, $n_{to}/t_c$,
is in rough equilibrium with their death rate, $n_S/t_S$.  
Assume further that the bolometric luminosity of a
star is simply $L \sim (M_*c^2/t_*)$.  Then $n_S = n_{to} (t_S/t_c)$,
and the ratio of bolometric luminosities emitted in steady state would be
\begin{equation}
\label{eq:lbolratio}
{L_S\over L_{to}} =\left({M_S \over M_{to}}\right) \left({t_{to}\over t_c}\right)
            \sim 2 \times \left({t_{to}} \over {2 \times 10^{10}} {\rm yr}\right).
\end{equation}
If the M32 central stellar population is old, i.e., $t_{to} \sim 10^{10}$ yr,
the flux ratio is of order unity, and the predicted fraction of emitted blue
flux would exceed that from P2 in M31 by a 5$\times$ (cf. $\S\ref{sec:dm31}$.
In fact, the color gradient discussion in $\S$\ref{sec:om32} failed to find any
blueing in the central pixels of M32, limiting it
to at most 15\%\ of that seen in M31.  

Some easing of this dilemma would
occur if the stellar population were young.  Estimates based on
H$\beta$ put the light-weighted central stellar age near $2 \times 10^9$
yr (\markcite{gon}Gonzalez 1993, \markcite{trag}Trager et al.\ 1998),
which would reduce the predicted blueing effect by a 5$\times$. 
However, the predicted blueing would still be comparable to that seen in M31,
as its estimated age is also rather young
($5 \times 10^9$ yr, Gonzalez (unpublished))
and the ratio $t_{to}/t_S$ is similar in the two galaxies.
Furthermore, the bulk of the population could be older than
the lightweighted age, and we have also neglected the extra collisions 
that would occur during the post-main sequence phases.
On the other hand, collision speeds may be high enough to prevent
efficient formation of blue stragglers (see $\S\ref{sec:dm31}$).
In sum, the stubborn refusal of
M32 to reveal any departures from utter normalcy at small radii may be 
stressing simple theories of its central structure, a problem that deserves
closer scrutiny with realistic multi-component evolving dynamical models.

\subsection{M33\label{sec:dm33}}
 
The key to the M33 nucleus may be that, unlike M32, it lacks a massive central
black hole to stabilize its structure.
\markcite{km}Kormendy \& McClure (1993) have emphasized their impressively
low upper limit of only $M_\bullet<5\times10^4M_\odot$
for a central black hole in M33;
with our smaller limit on the core radius,
we can reduce this further to $M_\bullet<2\times10^4M_\odot.$\@
With a central velocity dispersion of only $\sigma=21$ km s$^{-1}$ but
a central density nearly as high as that in M32, the stellar
collisions and relaxation, in particular, will be 
much more important for M33.
Stellar collisions indeed may be the explanation
for both the relatively blue color of the M33 nucleus and
its apparent population of young stars (\markcite{vdb}van den Bergh 1991).

\markcite{hern}Hernquist, Hut, \& Kormendy (1991) emphasized the short
central relaxation time in the M33 nucleus and the likelihood
that it has undergone core collapse, based on a preliminary
analysis of the \markcite{km}Kormendy \& McClure (1993) observations.
The present observations imply an even shorter relaxation time
than estimated in either work;
from equation (\ref{eq:trelax}) and assuming
$\sigma=r_c\sqrt{4\pi G\rho_0/9} = 21$ km s$^{-1}$ for M33,
\begin{equation}
\label{eq:trm33}
t_r=3\times10^6\left({r_c\over0.13~{\rm pc}}\right)^3
\left({\rho_0\over2\times10^6M_\odot~{\rm pc^{-3}}}\right)^{1/2}
\left({10\over\ln(0.4N)}\right){\rm yr}.
\end{equation}
The implied core-collapse time is only $t_{cc}\approx1\times10^9$ yr,
taking $t_{cc}\approx330t_r$ (\markcite{cohn}Cohn 1980);
it is thus highly likely that the nucleus has already undergone
core collapse.

\markcite{hern}Hernquist et al.\ (1991) were pessimistic that
the blue color of the nucleus could be explained via blue stragglers
generated in stellar mergers but did not present a quantitative argument.
\markcite{km}Kormendy \& McClure (1993), in contrast, were
intrigued that stellar collisions might indeed be important.
Simple calculations of the central stellar collision time scale suggest
that collisions are highly likely over the age of the nucleus.
For the case with $\sigma<<GM/r_\ast,$
equation (\ref{eq:tcoll}) reduces to
$t_c\approx\left[8\sqrt{\pi}G\rho r_p/\sigma\right]^{-1}.$\@
For the M33 nuclear parameters, this gives
\begin{equation}
\label{eq:tc}
t_c=7\times10^9\left({\rho\over 2\times10^6M_\odot~{\rm pc^{-3}}}\right)^{-1}
\left({\sigma\over 21~{\rm km~s^{-1}}}\right)~{\rm yr}.
\end{equation}
In other words, even for an intermediate-aged nucleus, the probability
that a solar-type star will have experienced a collision at the center
of M33 approaches unity.

If we assume that the present turnoff of the parent old population in M33
is at $\sim1M_\odot,$ then the typical $2M_\odot$ blue-straggler collision
product will resemble an A5 V star.
Explaining the larger $B-R$ versus $V-I$ color gradient requires only that stars
added to the center of the nucleus be of spectral type earlier
than F5 or so.  For an A5 star with $B-R\approx0.3,$ and $V-I\approx0.2,$
a fractional contribution to the total nuclear light of $\sim30\%$
is sufficient to explain both central color gradients;
a $\sim20\%$ contribution of A0 stars or a $\sim40\%$ contribution
of F0 stars would also work.

Due to the low mass-to-light ratios of A-type stars, the implied
mass contributions are modest and are well within the predicted mass
fraction of collision products.  For an assumed $M/L_V\approx2.5$
for the old nucleus population (see below) and an A5 luminosity of $16L_\odot$
at V, the implied mass-fraction of blue stragglers is only $\sim1\times10^{-2}.$\@
Allowing for the finite lifetime of an A5 star, $t\sim1.5\times10^9$ yr,
compared to the nominal $7\times10^9$ yr collision time still implies
that the blue population required is only $\sim6\%$ of that predicted by the
naive application of equation (\ref{eq:tc}).
Explaining the blue population with somewhat more
massive and hotter collision products, such as A0 stars, would imply
a still more modest collision efficiency.
Even a population of blue stragglers dominated by F0 stars still fits
within the naive predictions of the collision rate by an order of magnitude.

The sharp radial drop in stellar density in the outer nucleus means that
collisions will be effective only at the very center.  The collision rate
calculated above is for $r<0\asec03;$ at $r\sim0\asec1,$ the
rate is lower by an order of magnitude.
The strong dependence of collision rate with radius might explain
both the scale and onset of the central nuclear color gradients.
The typical stellar orbit in a potential similar to that of the nucleus
is likely to have a factor of three ratio between its pericenter
and apocenter distances (assuming an isotropic
velocity ellipsoid).  Any collision products generated within
$r<0\asec03$ are likely to have significant diffusion out to
$r\sim0\asec1$ but have little presence at still larger radii,
in qualitative agreement with the color gradients observed.
 
A qualitative argument suggests that the above collision rate may indeed
be somewhat overestimated.  If the M33 nucleus has undergone relaxation
and collapse as suggested by \markcite{hern}Hernquist, Hut, \& Kormendy (1991),
then significant mass segregation may have taken place at its
center, suggesting that the modest velocity dispersion measured by
\markcite{km}Kormendy \& McClure (1993) may be more representative of the
star {\it light} rather than mass.
The massive blue stragglers should be in equi-partition with the older stars,
suggesting that the true velocity dispersion, and hence collision time scale,
is higher.

A bias of this sort may in fact be suggested by the low
$M/L_V\approx0.4$ calculated by \markcite{km}Kormendy \& McClure (1993).
Models presented by \markcite{wor}Worthey (1994) show that the old nucleus
population will most likely have $M/L_V>2;$ a conclusion that
appears robust over a wide range of metallicities provided that
the bulk of the population is at least 5 Gyr old,
an assumption supported by the spectral synthesis work
of \markcite{sch}Schmidt, Bica, \& Alloin (1990) discussed below.
The low $M/L$ blue-straggler population will dilute this of course,
but not by the factor of five required;
the above model with 30\%\ of the light in A5 stragglers, for example,
has $M/L_V=1.8.$\@
To obtain an apparent $M/L_V=0.4$ would require measuring a $\sigma$
that is $\sim2\times$ too low, not unreasonable if much of the light
comes from A stars with masses $\sim4\times$ that of the typical star.
Regardless, such arguments imply that $t_c$ at best is only $\sim2\times$
longer than a naive application of equation (\ref{eq:tc}) might suggest.
This does not change the conclusion that collisions are important,
as the calculation still has a factor of $\sim5$ to spare.

The possibility that collisions are responsible for the young population
in M33 contrasts with the traditional conclusion that the M33
nucleus is a site of repeated episodes of star formation since
its initial formation (see \markcite{vdb}van den Bergh 1991).
Given the inherent freedom in appealing to episodic star formation,
we cannot argue that the
collision scenario is by any means favored, and indeed it may have difficulty
accounting for the UV spectral energy distribution of the nucleus.
\markcite{sch}Schmidt, Bica, \& Alloin (1990) use spectral synthesis to conclude
that a $\sim10^9$ yr population contributes $\sim30\%$ of the nuclear
light, consistent with the collision model, but they also
argue that a significant contribution from a very young $10^7$ yr
component is also present.  \markcite{mass}Massey et al.\ (1996) 
also infer a hot young component in the nucleus from far-UV photometry.  
Under the pure collision scenario, such hot components
would have to be stars produced in second-generation mergers
of the initial blue stragglers, or perhaps stars formed from gas
left over from mass-loss in the collisions.
Such effects can only be evaluated by full models of the nucleus
rather than the simple arguments presented here.
We only argue here merely point out 
that, unless the collision time has been severely underestimated,
collisions should have readily observable effects on the gross properties
of the M33 nucleus.

\section{Toward a Unified Picture\label{sec:conclude}}

We conclude with an attempt to place the Local Group nuclei into
a common evolutionary scenario.
As the nearest examples of nuclei in giant galaxies,
they may provide a view of phenomena in more distant nuclei, including AGNs.
For example, the steep density profiles of M31 and M32 place them in the
``power-law" category of central density profiles according to 
\markcite{l95}Lauer et al.\ (1995).
Understanding the origin of M31 and M32
may thus illuminate the origin of all power-law profiles, which
typify small ellipticals and bulges generally;
``power-law" galaxies may also serve as the building blocks of the shallower
core-type profiles of massive ellipticals (Faber et al.\ 1997).  

The Introduction strongly emphasized the distinct properties of 
these three nuclei;
in actual fact their inner density profiles prove to be broadly similar,
with central densities $\sim 10^{6.5 \pm 0.3}M_\odot~{\rm pc}^{-3}$ at 0.1 pc 
and radial slopes $\rho \sim r^{-1.5 \pm 0.5}$.
However, central dispersions differ enormously, with M31 and M32
in the range 200-800 km/s while M33 has a much lower $\sigma=21$ km s$^{-1}$.  
This difference of course reflects the putative presence of black holes;
M31 and M32 have estimated black hole masses in the
range $3\times 10^{6-7}M_\odot$,
while a black hole is undetectable in M33 at the level $<2\times10^4M_\odot.$\@

This tentative division into two types according to black hole content is
heightened by adding the fourth large galaxy in the Local Group, the Milky Way.
The Milky Way nucleus combines 
the properties of M32$\pm$1 in interesting ways:  it has a stellar density
profile that is rather similar to the other three, with a peak central density
at 0.1 pc $\sim 2 \times 10^6 M_\odot~{\rm pc}^{-3}$, falling off radially beyond
0.3 pc as $r^{-1.8}$ (\markcite{hall}Haller et al.\ 1996).  
Like M31 and M32, it has high central velocity dispersion
$\sigma(0.1~{\rm pc})\sim 200$ km s$^{-1}$ and an estimated central black hole
mass similar to M32 of $2 \times 10^6M_\odot$
(\markcite{hall}Haller et al.\ 1996).
However, it also has a young central cluster of blue stars (see review by
\markcite{mor}Morris \& Serabyn 1996), and in this regard it more resembles M33.
Significantly --- unlike any of the others --- the Milky Way is barred,
which strongly perturbs the central gas dynamics and appears to create the
conditions for a steady inflow of gas into the central $\sim 5$ pc at a rate of
roughly 0.03-0.05 $M_\odot$ yr$^{-1}$ (\markcite{mor}Morris \& Serabyn 1996).  
Bar aside, the Milky Way seems to fit well with M31 and M32.
Indeed, from the standpoint of central structure, the Milky Way
looks like a barred spiral galaxy with a transplanted M32 spheroid.
For purposes of comparison, we therefore lump it with M32 and M31.

As a group, it may be significant that the stellar collision time scales
for all four Local Group galaxies hover at or below values that would
have barely detectable dynamical effects over a Hubble time.
This suggests that the systems have evolved via similar
paths to this state, or will do so eventually.
One such scenario has been outlined by
\markcite{mcd}Murphy, Cohn, \& Durisen (1991, MCD),
who have modeled the evolution of
compact star clusters in the vicinity of seed black holes, including the
effects of black hole growth and stellar relaxation, stellar mass
loss, and stellar collisions (but not mass ejection by supernovae).  
The inner density profiles and black hole masses of their Models 3 and
4 at an age of $\sim10$ Gyr are qualitatively consistent with 
M31, M32, and the Milky Way.
Interestingly, the initial central density
profiles in all MCD models start out much {\it lower} and {\it flatter}
within 1 pc than the evolved
profiles ---  the dense, steep profiles seen at $\sim10$ Gyr 
are due to the combined effects of 
stellar relaxation, loss-cone depletion, and adiabatic
pulling by the growing black hole.
Early evolution of the stellar profile is
rapid owing to the initially low stellar velocity dispersion, which induces
short stellar relaxation and collision times.
As the black hole grows, however, the stellar velocity dispersion
rises and the stellar population gets ``stiffer;''
relaxation shuts down and collisions become less frequent.
This is the same stabilizing by the black hole that was invoked by
\markcite{l92}Lauer et al.\ (1992) to postpone
rapid dynamical evolution in M32.
The black hole continues to grow at late times via stellar mass loss and
loss-cone capture, but only slowly, and the systems after 10 Gyr appear
to be in near-steady state, like M31, M32, and the Milky Way.

An important question is why the initial conditions of M33's
nucleus set its evolution on a different path from the that of the others.
A possibility is that the M33 galaxy lacks a massive spheroidal
component while the other three galaxies have one.
If initial central density is the key to subsequent evolution, as the MCD models 
suggest, then the presence of a surrounding spheroid, and consequent
increase in stellar mass loss and/or merger-induced 
gaseous infall, may be needed to boost nuclear density to the
required level to grow a black hole quickly.
So far, all detected massive central black holes are in galaxies
with spheroids (\markcite{r98}Richstone et al.\ 1998)
save for one Seyfert nucleus in the bulgeless dwarf
Sdm galaxy NGC 4395 (\markcite{fil}Filippenko, Ho \& and Sargent 1993).
A major task for the {\it Hubble Space Telescope} is to search for
black holes in bulgeless galaxies.

The lack of a central black hole in the M33 nucleus leads
to its profoundly short central relaxation time, $t_r\approx3\times10^6$ yr,
several orders of magnitude shorter than that
for M32 or M31 (see Table \ref{tab:time}).
We have advocated the short central {\it collision} time in M33 as a possible
explanation for its blue center;
however, the lack of any apparent collision effects in M32, with
a central collision time only $\sim3\times$ longer serves as a strong
caveat that this hypothesis as simply stated may be incomplete.
The short relaxation time implies a strong likelihood that the M33
nucleus has core-collapsed.
The collision probability increases strongly during collapse,
suggesting that the blue-straggler population was formed during the
collapse, rather than under any sort of quasi-static conditions.
Indeed, the short relaxation time of the nucleus already implies that any
naive calculations based on the presumption that its structure has not
evolved over the age of its stellar population are likely to be in error.
In contrast, the long relaxation times in M31 and M32 argue for
little recent structural evolution, suggesting that the naive
collision calculations may be closer to the truth in these two nuclei.

The longer native collision time in M31 compared to M32 makes the
presence of a central blue excess there but not in M32 a compelling problem.
Because the blue region is resolved, it must be produced mainly by starlight.
This is interesting because all conceivable types of hot stars ---
blue stragglers, hot horizontal branch stars, young upper main-sequence
stars --- are {\it short-lived}.
Hence, understanding the origin of this blue
flux sets constraints on current or very recent events in M31.
Since the main-sequence stellar collision time is 
longer in M31 than in M32 (see $\S\ref{sec:dm31}$ and $\S\ref{sec:dm32}$),
we must appeal to some other difference between the two nuclei.
We consider some possibilities in turn:

\begin{enumerate}
\item The velocity dispersion near the black hole is $\sim3\times$ larger
at a given radius in M31 than in M32, so stellar collisions are more violent.
As noted in $\S$\ref{sec:dm31},
this works {\it against} the blue-straggler hypothesis 
for the blue excess in M31, as blue stragglers are less likely to
coalesce at these high speeds.
The higher overall velocity dispersion in M31 thus does
not favor the formation of hot stars, at least not by mergers.

\item The estimated black hole mass in M31 is $\sim10\times$ larger than
that in M32, and the maximum Eddington radiation limit
is larger by the same factor.
If both black holes radiated occasionally at their respective limits,
the black hole in M31 would evaporate the envelopes
of giants to a larger radius, revealing more bare hot cores.  
According to \markcite{tout}Tout et al.\ (1989),
a radiation bath of $10^4$ K causes K giants to lose their envelopes.
However, at the M31 black hole Eddington limit of $10^{12}L_\odot$,
the effective temperature of the radiation field falls to this level
at only 1000 AU, $\sim40\times$ smaller than region of the blue excess (0.2 pc).
This model would also require an {\it ad hoc} recent
maximal burst of AGN activity in M31 within the last few $10^8$ yr.

\item As noted in $\S\ref{sec:dm31}$ and $\S\ref{sec:dm32}$,
the central stellar population may be younger in M32 than in
M31, ($2 \times 10^9$ yr vs. $5 \times 10^9$ yr).  
By Equation (\ref{eq:lbolratio}), this
age difference merely nullifies the
difference in the collision times, which also differ by a
similar factor; it does not explain why
the fractional blue excess is so much {\it larger} in M31 than in M32.

\item A cold disk of stars in elliptical orbits is required in the
\markcite{trem}Tremaine (1995) model for the M31 double nucleus, 
while no such disk is seen in M32.
Suppose that some fraction $f$ of stars within 0.2 pc in M31 are orbiting
in a cold disk with $\sigma_{disk} = 0.05 v_{rot}$,
which is typical of young disk stars in the Milky Way.
At 0.1 pc,  the random velocity dispersion in 
the disk, $\sigma_{disk}$, is $\sim 60$ km s$^{-1}$, $\sim120\times$
colder than the isotropic velocity dispersion of 700 km s$^{-1}$.\@
The focusing enhancement of the collision cross section over the geometrical
value is the Safronov number (\ref{eq:tcoll}) given by
$\Theta = (V_\ast^{esc}/2 \sigma_{disk})^2$,
where $V_\ast^{esc}$ is the escape velocity from
the stellar surface (600 km s$^{-1}$
for the sun), and $\sigma_{disk}$ is the 1-d velocity dispersion of the disk.
In addition, the effective density of the disk stars is also increased
by the factor $\sim \sigma/\sigma_{disk}$ as they are compressed into the disk.
Under these assumptions, the new collision time is reduced by the factor
$\sim\Theta^{-1}$, which for $\sigma_{disk}=60$ km s$^{-1}$, is a factor of 25.
The new collision time is $2.4 \times 10^9$ yr,
and thus essentially all disk stars would suffer a low-speed
collision, causing them to merge.
Roughly a fraction $f$ of the central light would be
emitted with the effective temperature of blue stragglers, and thus 
$f \sim 0.2$ is needed to match the magnitude of the blue excess.
Though promising, this model still does not
reach the high color temperature of $>10^4$ K that seems to be needed
to match the spectral-energy distribution of the blue excess.

\item M31 possesses neutral gas like the Milky Way
--- unlike M32, which is barren.
The region within 1 pc of the Milky Way black hole appears to be a site of
episodic star formation, with a mass of stars
formed in the last few million years of about $10^4M_\odot$ 
(\markcite{mor}Morris \& Serabyn 1996).
In M31, the color temperature of the blue excess
indicates a main-sequence turnoff mass near A0 and therefore an age
of about $10^9$ yr if modeled as a single-burst population.
The blue-excess V luminosity of $1.7 \times 10^4L_\odot$
and resultant $M/L_V \sim 1$ (\markcite{bru}Bruzual \& Charlot 1993) 
imply a mass of $10^4M_\odot$, similar to the Milky Way central cluster.
The radii of the clusters are also similar, a few tenths of a parsec. 
If allowed to age undisturbed, the young population centered on the Milky
Way black hole would evidently look rather like the blue excess population
in M31 in about 1 Gyr.  
This explanation begs the question of forming stars in such a region,
where the random velocities exceed several $10^2$
km s$^{-1}$ and the shear is intense,
but the Milky Way appears to have solved that problem somehow, 
albeit with velocities about a factor of three smaller.
It is also true that the center of M31 is relatively empty of neutral gas now,
but an intermittent bar instability might bring gas to the center
on periods of a few Gyr.
It is intriguing that, in this model,
the dynamical structure of M31 might have looked quite different in the past.
\end{enumerate}

None of these proposals for explaining the blue excess population in
M31 is without drawbacks.  However, all illuminate the importance of carrying
studies of galactic nuclei to the next stage, where all significant processes
are modeled and related nuclei are compared and contrasted.  
We hope that the emerging body of high-resolution data on Local Group galaxies
will stimulate modelers to match in detail the evolutionary history of these
nearby Rosetta Stone objects.

\acknowledgments

We wish to thank Doug Lin as well as our many collaborators
on the Nuker team for useful discussions.
This research was conducted by the WFPC-1 Investigation Definition Team,
supported in part by NASA Grant No. NAS5-1661.

\clearpage

\begin{deluxetable}{clclcl}
\tablecolumns{6}
\tablewidth{0pt}
\tablecaption{Observational Summary}
\tablehead{
\colhead{Galaxy}& \colhead{Filter}& \colhead{Date}& \colhead{Exposure (s)}&
\colhead{Peak (e$^-$)}& \colhead{Notes}}
\startdata
M31&F160BW&06/18/95&$6\times2000$&\dots&No clear signal\nl
   &F300W&06/19/95&$6\times500$&$2.1\times10^3$& \nl
   &F555W&06/19/95&$4\times300$&$9.0\times10^4$&Dither sequence, gain=15\nl
   &     &06/19/95&$2\times500$&\dots&Saturated in center\nl
   &F814W&06/19/95&$4\times300$&$1.4\times10^5$&Dither sequence, gain=15\nl
   &     &06/19/95&$2\times500$&\dots&Saturated in center\nl
   &     &       &           &               & \nl
M32&F555W&12/26/94&$4\times26$&$4.0\times10^4$& \nl
   &F814W&12/26/94&$4\times26$&$5.0\times10^4$& \nl
   &     &       &           &               & \nl
M33&F555W&01/19/95&$4\times20$&$2.4\times10^4$& \nl
   &     &01/19/95&$2\times500$&\dots&Saturated in center\nl
   &F814W&01/19/95&$6\times10$&$1.6\times10^4$& \nl
\enddata
\label{tab:obs}
\end{deluxetable}
\begin{deluxetable}{lcccc}
\tablecolumns{5}
\tablewidth{0pt}
\tablecaption{M31 Nuclear Colors}
\tablehead{
\colhead{Color}& \colhead{P1}& \colhead{P2}& \colhead{Anti-P1}&
\colhead{$1''<r<5''$}}
\startdata
$V-I$&$1.41\pm0.01$&$1.28\pm0.03$&$1.38\pm0.01$&$1.34\pm0.01$\nl
$F300-V$&$2.29\pm0.02$&$1.29\pm0.03$&$2.18\pm0.02$&$2.39\pm0.02$\nl
\enddata
\label{tab:m31nuc}
\tablecomments{The P1 and anti-P1 colors were measured in
$0\asec4\times0\asec4$ boxes centered on P1 and displaced $0\asec4$
from P2 in the anti-P1 direction, respectively.
The P2 colors were measured in a smaller $0\asec14\times0\asec14$
box to isolate the UV-bright portion of the nucleus.
The last column represents the bulge outside the nucleus;
the radial range is distance from P2.}
\end{deluxetable}

\clearpage

\begin{deluxetable}{rlcclc}
\tablecolumns{6}
\tablewidth{0pt}
\tablecaption{M31 Deconvolved Surface Photometry}
\tablehead{
\colhead{$a$}& \colhead{$\mu_V$}& \colhead{$F300-V$}& \colhead{$V-I$}& \colhead{PA}&
\colhead{$\epsilon$}}
\startdata
 0.0114&$13.235\pm0.050$&$0.763\pm0.070$&$1.169\pm0.070$&$\cdots$&$\cdots$\nl
 0.0227&$13.340\pm0.004$&$0.658\pm0.004$&$1.240\pm0.005$&$ 60.9\pm  8.4$&$0.254\pm0.037$\nl
 0.0455&$13.529\pm0.007$&$0.655\pm0.009$&$1.357\pm0.009$&$ 60.9\pm  3.1$&$0.320\pm0.017$\nl
 0.0683&$13.603\pm0.020$&$0.921\pm0.023$&$1.383\pm0.025$&$ 60.9\pm 10.4$&$0.484\pm0.087$\nl
 0.0910&$13.621\pm0.018$&$1.143\pm0.025$&$1.377\pm0.025$&$ 64.6\pm 25.1$&$0.613\pm0.260$\nl
 0.1137&$13.643\pm0.014$&$1.246\pm0.026$&$1.371\pm0.020$&$ 60.4\pm 10.8$&$0.671\pm0.126$\nl
 0.1365&$13.683\pm0.012$&$1.308\pm0.040$&$1.374\pm0.016$&$ 57.8\pm  4.2$&$0.671\pm0.049$\nl
 0.1592&$13.709\pm0.009$&$1.356\pm0.025$&$1.374\pm0.013$&$ 56.8\pm  4.1$&$0.671\pm0.048$\nl
 0.1820&$13.733\pm0.008$&$1.397\pm0.030$&$1.394\pm0.013$&$ 54.1\pm  3.5$&$0.653\pm0.040$\nl
 0.2048&$13.748\pm0.007$&$1.484\pm0.023$&$1.372\pm0.010$&$ 48.7\pm  4.5$&$0.634\pm0.049$\nl
 0.2275&$13.790\pm0.006$&$1.533\pm0.019$&$1.404\pm0.008$&$ 46.8\pm  1.3$&$0.588\pm0.013$\nl
 0.2828&$13.794\pm0.007$&$1.716\pm0.022$&$1.363\pm0.009$&$ 46.7\pm  3.4$&$0.578\pm0.022$\nl
 0.3327&$13.812\pm0.007$&$1.833\pm0.017$&$1.341\pm0.009$&$ 56.7\pm  6.6$&$0.577\pm0.014$\nl
 0.3915&$13.924\pm0.007$&$1.857\pm0.020$&$1.392\pm0.009$&$ 53.9\pm  4.9$&$0.500\pm0.017$\nl
 0.4605&$13.962\pm0.006$&$1.870\pm0.018$&$1.384\pm0.008$&$ 57.3\pm  1.4$&$0.520\pm0.027$\nl
 0.5418&$14.014\pm0.006$&$1.838\pm0.023$&$1.322\pm0.007$&$ 62.4\pm  0.5$&$0.536\pm0.007$\nl
 0.6374&$14.227\pm0.006$&$1.766\pm0.021$&$1.354\pm0.007$&$ 65.0\pm  0.6$&$0.402\pm0.008$\nl
 0.7499&$14.330\pm0.005$&$1.826\pm0.018$&$1.338\pm0.006$&$ 62.7\pm  0.5$&$0.410\pm0.006$\nl
 0.8822&$14.497\pm0.005$&$1.914\pm0.016$&$1.380\pm0.006$&$ 62.7\pm  0.4$&$0.405\pm0.005$\nl
 1.0379&$14.695\pm0.005$&$1.934\pm0.014$&$1.296\pm0.006$&$ 62.4\pm  0.4$&$0.361\pm0.005$\nl
 1.2211&$14.893\pm0.004$&$1.886\pm0.011$&$1.356\pm0.005$&$ 60.5\pm  0.4$&$0.332\pm0.004$\nl
 1.4366&$15.108\pm0.004$&$1.847\pm0.013$&$1.337\pm0.005$&$ 55.1\pm  0.5$&$0.270\pm0.004$\nl
 1.6901&$15.212\pm0.003$&$1.897\pm0.012$&$1.324\pm0.004$&$ 58.0\pm  0.4$&$0.308\pm0.004$\nl
 1.9883&$15.356\pm0.002$&$1.976\pm0.009$&$1.334\pm0.003$&$ 54.7\pm  0.4$&$0.294\pm0.003$\nl
 2.0759&$15.429\pm0.002$&$1.970\pm0.011$&$1.386\pm0.003$&$ 55.3\pm  0.5$&$0.223\pm0.003$\nl
 2.4422&$15.523\pm0.001$&$1.954\pm0.009$&$1.372\pm0.002$&$ 58.2\pm  0.3$&$0.252\pm0.003$\nl
 2.8732&$15.653\pm0.001$&$2.042\pm0.009$&$1.336\pm0.001$&$ 55.6\pm  0.3$&$0.219\pm0.003$\nl
 3.3802&$15.768\pm0.001$&$1.981\pm0.009$&$1.344\pm0.001$&$ 55.0\pm  0.6$&$0.137\pm0.003$\nl
 3.9767&$15.861\pm0.001$&$2.026\pm0.007$&$1.348\pm0.001$&$ 48.7\pm  0.6$&$0.135\pm0.002$\nl
 4.6785&$15.951\pm0.001$&$2.024\pm0.006$&$1.338\pm0.001$&$ 58.8\pm  0.5$&$0.133\pm0.002$\nl
 5.5041&$16.026\pm0.001$&$\cdots$&$\cdots$&$ 49.6\pm  0.7$&$0.107\pm0.002$\nl
 6.4754&$16.092\pm0.001$&$\cdots$&$\cdots$&$ 47.3\pm  0.4$&$0.154\pm0.002$\nl
 7.6181&$16.151\pm0.001$&$\cdots$&$\cdots$&$ 47.8\pm  0.3$&$0.182\pm0.002$\nl
 8.9624&$16.219\pm0.001$&$\cdots$&$\cdots$&$ 42.2\pm  0.2$&$0.203\pm0.001$\nl
10.5440&$16.306\pm0.000$&$\cdots$&$\cdots$&$ 40.1\pm  0.2$&$0.170\pm0.001$\nl
\enddata
\label{tab:m31surf}
\tablecomments{V-band deconvolved surface brightness profile
is given as a function of isophote semimajor axis length in arcseconds.
The area dominated by P1 has been excluded from the fit.}
\end{deluxetable}

\clearpage

\begin{deluxetable}{rlclcr}
\tablecolumns{6}
\tablewidth{0pt}
\tablecaption{M32 Deconvolved Surface Photometry}
\tablehead{
\colhead{$a$}& \colhead{$\mu_V$}& \colhead{$V-I$}& \colhead{PA}&
\colhead{$\epsilon$}&\colhead{$100\times A_4$}}
\startdata
 0.0161&$10.959\pm0.070$&$1.174\pm0.090$&$\cdots$&$\cdots$&$\cdots$\nl
 0.0455&$11.228\pm0.008$&$1.216\pm0.009$&$160.3\pm10.5$&$0.146\pm0.027$&$-0.7\pm 1.5$\nl
 0.0910&$11.642\pm0.004$&$1.247\pm0.006$&$160.3\pm 1.1$&$0.230\pm0.004$&$-2.9\pm 0.8$\nl
 0.1365&$11.891\pm0.007$&$1.239\pm0.017$&$160.3\pm 1.8$&$0.285\pm0.009$&$ 0.1\pm 1.2$\nl
 0.1820&$12.094\pm0.008$&$1.255\pm0.012$&$160.3\pm 1.7$&$0.286\pm0.009$&$ 0.0\pm 1.1$\nl
 0.2275&$12.235\pm0.008$&$1.220\pm0.013$&$160.3\pm 2.1$&$0.285\pm0.010$&$ 2.4\pm 1.1$\nl
 0.2730&$12.425\pm0.007$&$1.223\pm0.013$&$160.3\pm 1.3$&$0.259\pm0.006$&$-0.2\pm 0.9$\nl
 0.3185&$12.568\pm0.008$&$1.230\pm0.013$&$159.8\pm 1.8$&$0.247\pm0.008$&$-0.1\pm 1.0$\nl
 0.3640&$12.686\pm0.010$&$1.241\pm0.018$&$159.7\pm 2.3$&$0.243\pm0.010$&$-0.7\pm 1.1$\nl
 0.4095&$12.787\pm0.007$&$1.255\pm0.014$&$159.7\pm 1.8$&$0.247\pm0.008$&$-1.5\pm 0.8$\nl
 0.4550&$12.898\pm0.008$&$1.246\pm0.014$&$159.7\pm 1.7$&$0.247\pm0.007$&$ 1.9\pm 0.9$\nl
 0.5005&$12.979\pm0.008$&$1.225\pm0.014$&$159.7\pm 2.0$&$0.255\pm0.009$&$ 2.9\pm 0.8$\nl
 0.5460&$13.076\pm0.008$&$1.251\pm0.014$&$159.5\pm 1.5$&$0.259\pm0.007$&$-0.6\pm 0.8$\nl
 0.5915&$13.164\pm0.008$&$1.263\pm0.012$&$159.7\pm 1.4$&$0.259\pm0.006$&$-0.5\pm 0.7$\nl
 0.6370&$13.250\pm0.007$&$1.264\pm0.011$&$160.0\pm 1.3$&$0.259\pm0.006$&$-1.4\pm 0.7$\nl
 0.6825&$13.354\pm0.004$&$1.258\pm0.008$&$160.0\pm 0.5$&$0.259\pm0.002$&$-0.6\pm 0.4$\nl
 0.7280&$13.426\pm0.006$&$1.250\pm0.010$&$160.0\pm 1.1$&$0.262\pm0.005$&$-1.4\pm 0.5$\nl
 0.7829&$13.509\pm0.011$&$1.217\pm0.014$&$160.3\pm 1.3$&$0.264\pm0.009$&$ 0.6\pm 0.4$\nl
 0.9211&$13.746\pm0.010$&$1.223\pm0.013$&$159.1\pm 1.1$&$0.281\pm0.008$&$-1.2\pm 0.7$\nl
 1.0836&$13.951\pm0.010$&$1.248\pm0.013$&$157.7\pm 0.9$&$0.290\pm0.008$&$-0.2\pm 0.3$\nl
 1.2748&$14.203\pm0.010$&$1.225\pm0.013$&$159.1\pm 0.9$&$0.288\pm0.007$&$-0.1\pm 0.5$\nl
 1.4998&$14.450\pm0.009$&$1.243\pm0.012$&$158.9\pm 0.8$&$0.282\pm0.007$&$-0.3\pm 0.1$\nl
 1.7645&$14.684\pm0.009$&$1.239\pm0.011$&$160.5\pm 0.8$&$0.293\pm0.007$&$ 0.1\pm 0.2$\nl
 2.0759&$14.906\pm0.008$&$1.234\pm0.011$&$159.0\pm 0.7$&$0.301\pm0.007$&$-0.1\pm 0.2$\nl
 2.4422&$15.142\pm0.008$&$1.223\pm0.010$&$159.6\pm 0.8$&$0.284\pm0.007$&$-0.1\pm 0.3$\nl
 2.8732&$15.351\pm0.007$&$1.225\pm0.009$&$157.9\pm 0.7$&$0.293\pm0.006$&$ 0.0\pm 0.2$\nl
 3.3802&$15.580\pm0.007$&$1.233\pm0.009$&$159.9\pm 0.7$&$0.276\pm0.006$&$ 0.1\pm 0.3$\nl
 3.9767&$15.785\pm0.006$&$1.244\pm0.008$&$159.1\pm 0.7$&$0.279\pm0.006$&$-0.2\pm 0.1$\nl
 4.6785&$15.997\pm0.006$&$1.222\pm0.008$&$159.2\pm 0.7$&$0.269\pm0.005$&$ 0.6\pm 0.3$\nl
 5.5041&$16.194\pm0.005$&$1.238\pm0.007$&$158.2\pm 0.6$&$0.283\pm0.005$&$ 0.2\pm 0.2$\nl
 6.4754&$16.408\pm0.005$&$1.219\pm0.006$&$158.6\pm 0.5$&$0.290\pm0.004$&$ 0.0\pm 0.2$\nl
 7.6181&$16.672\pm0.005$&$1.239\pm0.006$&$161.6\pm 0.4$&$0.292\pm0.004$&$ 0.2\pm 0.2$\nl
 8.9624&$16.898\pm0.005$&$1.220\pm0.006$&$160.2\pm 0.4$&$0.300\pm0.004$&$\cdots$\nl
10.5440&$17.175\pm0.007$&$1.243\pm0.009$&$158.8\pm 0.5$&$0.284\pm0.005$&$\cdots$\nl
\enddata
\label{tab:m32surf}
\tablecomments{V-band deconvolved surface brightness profile
is given as a function of isophote semimajor axis length in arcseconds.}
\end{deluxetable}

\clearpage

\begin{deluxetable}{rllrc}
\tablecolumns{6}
\tablewidth{0pt}
\tablecaption{M33 Deconvolved Surface Photometry}
\tablehead{
\colhead{$a$}& \colhead{$\mu_V$}& \colhead{$V-I$}& \colhead{PA}&
\colhead{$\epsilon$}}
\startdata
 0.0161&$10.887\pm0.070$&$0.925\pm0.090$&$\cdots$&$\cdots$\nl
 0.0455&$11.477\pm0.025$&$0.872\pm0.031$&$  6.2\pm38.0$&$0.052\pm0.032$\nl
 0.0910&$12.560\pm0.029$&$0.835\pm0.038$&$  6.2\pm12.2$&$0.038\pm0.008$\nl
 0.1365&$13.089\pm0.062$&$0.883\pm0.075$&$  6.2\pm11.2$&$0.159\pm0.031$\nl
 0.1820&$13.628\pm0.022$&$0.940\pm0.039$&$171.0\pm\phantom{0}4.0$&$0.114\pm0.008$\nl
 0.2275&$14.032\pm0.027$&$1.004\pm0.034$&$  9.6\pm\phantom{0}5.6$&$0.116\pm0.011$\nl
 0.2730&$14.353\pm0.028$&$1.003\pm0.046$&$ 17.8\pm\phantom{0}6.4$&$0.116\pm0.013$\nl
 0.3185&$14.645\pm0.027$&$1.059\pm0.057$&$ 21.7\pm\phantom{0}5.4$&$0.124\pm0.012$\nl
 0.3640&$14.942\pm0.028$&$1.034\pm0.047$&$ 21.7\pm\phantom{0}4.4$&$0.137\pm0.010$\nl
 0.4095&$15.236\pm0.025$&$1.032\pm0.044$&$ 22.0\pm\phantom{0}3.4$&$0.142\pm0.008$\nl
 0.4550&$15.394\pm0.029$&$1.006\pm0.047$&$ 23.0\pm\phantom{0}6.2$&$0.159\pm0.017$\nl
 0.5005&$15.609\pm0.033$&$1.049\pm0.053$&$ 22.7\pm\phantom{0}4.6$&$0.159\pm0.013$\nl
 0.5460&$15.839\pm0.035$&$1.030\pm0.050$&$ 23.0\pm\phantom{0}4.0$&$0.165\pm0.012$\nl
 0.5915&$16.020\pm0.035$&$0.909\pm0.051$&$ 23.0\pm\phantom{0}4.8$&$0.165\pm0.014$\nl
 0.6370&$16.120\pm0.033$&$1.003\pm0.050$&$ 22.7\pm\phantom{0}7.0$&$0.186\pm0.023$\nl
 0.6825&$16.199\pm0.030$&$1.005\pm0.045$&$ 23.0\pm\phantom{0}7.6$&$0.186\pm0.025$\nl
 0.7280&$16.365\pm0.030$&$1.084\pm0.048$&$ 22.7\pm\phantom{0}3.4$&$0.180\pm0.011$\nl
 0.7735&$16.473\pm0.053$&$1.011\pm0.075$&$ 20.2\pm\phantom{0}9.0$&$0.180\pm0.028$\nl
 0.8190&$16.634\pm0.028$&$0.995\pm0.056$&$ 20.2\pm\phantom{0}3.2$&$0.164\pm0.009$\nl
 0.8645&$16.796\pm0.031$&$0.993\pm0.057$&$ 20.2\pm\phantom{0}3.3$&$0.164\pm0.009$\nl
 0.9100&$16.969\pm0.037$&$0.920\pm0.070$&$  7.4\pm\phantom{0}3.5$&$0.164\pm0.010$\nl
 0.9555&$17.107\pm0.036$&$0.939\pm0.056$&$  7.4\pm\phantom{0}3.8$&$0.177\pm0.012$\nl
 1.0010&$17.130\pm0.037$&$0.979\pm0.059$&$  7.4\pm24.3$&$0.177\pm0.073$\nl
 1.0465&$17.037\pm0.096$&$0.766\pm0.110$&$179.8\pm16.0$&$0.177\pm0.049$\nl
 1.0920&$17.356\pm0.031$&$0.930\pm0.057$&$  0.1\pm\phantom{0}1.4$&$0.147\pm0.004$\nl
 1.1375&$17.430\pm0.037$&$0.957\pm0.064$&$  8.7\pm\phantom{0}6.6$&$0.169\pm0.020$\nl
\enddata
\label{tab:m33surf}
\tablecomments{V-band deconvolved surface brightness profile
is given as a function of isophote semimajor axis length in arcseconds.}
\end{deluxetable}

\clearpage

\begin{deluxetable}{ccccll}
\tablecolumns{6}
\tablewidth{0pt}
\tablecaption{Time Scales at 0.1 pc}
\tablehead{
\colhead{ }& \colhead{$\rho$}& \colhead{$\sigma$}& \colhead{ }&
\colhead{$t_r$}& \colhead{$t_c$} \\
\colhead{Galaxy}& \colhead{($M_\odot~{\rm pc}^{-3})$}& \colhead{(km s$^{-1}$)}&
\colhead{$\Theta$}& \colhead{(yr)}& \colhead{(yr)}}
\startdata
M31 ``Blue P2''&$1\times10^6$&720&0.2&$7\times10^{11}$&$6\times10^{10}$\nl
M31 ``Outer P2''&$4\times10^5$&720&0.2&$2\times10^{12}$&$2\times10^{11}$\nl
M32&$7\times10^6$&240&1.6&$3\times10^{9}$&$2\times10^{10}$\nl
M33&$2\times10^6$&21&220&$3\times10^{6}$&$7\times10^{9}$\nl
\enddata
\label{tab:time}
\tablecomments{Relaxation times and stellar collision times
are given for solar-type stars at 0.1 pc from the centers
of M$32\pm1;$ these are calculated from the local mass densities,
velocity dispersions, and Safronov numbers shown.
The two M31 entries are for densities implied by the blue P2 center
directly, or extrapolated inwards from the outer P2 cusp (see text).}
\end{deluxetable}

\end{document}